\RequirePackage{fix-cm}
\documentclass[twocolumn,epjc3]{svjour3}  
\smartqed  
\RequirePackage{mathptmx}
\RequirePackage{graphicx}
\RequirePackage{subfigure}
\RequirePackage{rotating}
\RequirePackage{pdflscape}
\RequirePackage{color}
\RequirePackage{lineno}
\RequirePackage{latexsym}
\RequirePackage{xspace}

\usepackage{amsmath,amsfonts,bm}
\usepackage{wasysym}
\usepackage{upgreek}
\newcommand{\onbb}{$0\nu\beta\beta$\xspace}
\newcommand{\nnbb}{$2\nu\beta\beta$\xspace}

\newcommand{\edw}{EDELWEISS\xspace}
\newcommand{\cuore}{CUORE\xspace}
\newcommand{\cupid}{CUPID\xspace}
\newcommand{\cupido}{CUPID-0\xspace}
\newcommand{\cupidmo}{CUPID-Mo\xspace}
\newcommand{\rly}{RLY\xspace}

\newcommand{\Qbb}{$Q_{\beta\beta}$\xspace}

\newcommand{\Mo}{$^{100}$Mo\xspace}
\newcommand{\lmo}{Li$_2$MoO$_4$\xspace}
\newcommand{\enrLMO}{Li$_{2}${}$^{100}$MoO$_4$\xspace}

\newcommand{\enrZS}{Zn$^{82}$Se\xspace}

\newcommand{\ntd}{NTD\xspace}

\newcommand{\TL}{$^{208}\mathrm{Tl}$\xspace}
\newcommand{\Co}{$^{60}\mathrm{Co}$\xspace}

\newcommand{\PO}{$^{210}\mathrm{Po}$\xspace}

\newcommand{\BI}{$^{214}\mathrm{Bi}$\xspace}
\newcommand{\PT}{$^{190}\mathrm{Pt}$\xspace}
\newcommand{\al}{$\alpha$\xspace}
\newcommand{\be}{$\beta$\xspace}
\newcommand{\ga}{$\gamma$\xspace}
\newcommand{\xr}{X-ray\xspace}
\newcommand{\ky}{kg$\times$yr\xspace}
\newcommand{\ckky}{counts/(keV$\times$kg$\times$yr)\xspace}

\journalname{Eur. Phys. J. C}
\begin{document}

\title{The \cupidmo experiment for neutrinoless double-beta decay: performance and prospects }



\author{
E.~Armengaud\thanksref{CEA-IRFU}\and
C.~Augier\thanksref{IPNL}\and
A.~S.~Barabash\thanksref{ITEP}\and
F.~Bellini\thanksref{Sapienza,INFN-Roma}\and
G.~Benato\thanksref{UCB}\and
A.~Beno\^{\i}t\thanksref{Neel}\and
M.~Beretta\thanksref{Milano,INFN-Milano}\and
L.~Berg\'e\thanksref{CSNSM}\and
J.~Billard\thanksref{IPNL}\and
Yu.~A.~Borovlev\thanksref{NIIC}\and
Ch.~Bourgeois\thanksref{LAL}\and
M.~Briere\thanksref{LAL}\and
V.~B.~Brudanin\thanksref{JINR}\and
P.~Camus\thanksref{Neel}\and
L.~Cardani\thanksref{INFN-Roma}\and
N.~Casali\thanksref{INFN-Roma}\and
A.~Cazes\thanksref{IPNL}\and
M.~Chapellier\thanksref{CSNSM}\and
F.~Charlieux\thanksref{IPNL}\and
M.~de~Combarieu\thanksref{CEA-IRAMIS}\and
I.~Dafinei\thanksref{INFN-Roma}\and
F.~A.~Danevich\thanksref{KINR}\and
M.~De~Jesus\thanksref{IPNL}\and
L.~Dumoulin\thanksref{CSNSM}\and
K.~Eitel\thanksref{KIT-IK}\and
E.~Elkhoury\thanksref{IPNL}\and
F.~Ferri\thanksref{CEA-IRFU}\and
B.~K.~Fujikawa\thanksref{LBNLNSD}\and
J.~Gascon\thanksref{IPNL}\and
L.~Gironi\thanksref{Milano,INFN-Milano}\and
A.~Giuliani\thanksref{e1,CSNSM,DISAT}\and
V.~D.~Grigorieva\thanksref{NIIC}\and
M.~Gros\thanksref{CEA-IRFU}\and
E.~Guerard\thanksref{LAL}\and
D.~L.~Helis\thanksref{CEA-IRFU}\and
H.~Z.~Huang\thanksref{Fudan}\and
R.~Huang\thanksref{UCB}\and
J.~Johnston\thanksref{MIT}\and
A.~Juillard\thanksref{IPNL}\and
H.~Khalife\thanksref{CSNSM}\and 
M.~Kleifges\thanksref{KIT-IPE}\and
V.~V.~Kobychev\thanksref{KINR}\and
Yu.~G.~Kolomensky\thanksref{UCB,LBNLPH}\and
S.I.~Konovalov\thanksref{ITEP}\and
A.~Leder\thanksref{MIT}\and
P.~Loaiza\thanksref{LAL}\and
L.~Ma\thanksref{Fudan}\and
E.~P.~Makarov\thanksref{NIIC}\and
P.~de~Marcillac\thanksref{CSNSM}\and
L.~Marini\thanksref{UCB, LBNLNSD, LNGS}\and
S.~Marnieros\thanksref{CSNSM}\and
X.-F.~Navick\thanksref{CEA-IRFU}\and
C.~Nones\thanksref{CEA-IRFU}\and
V.~Novati\thanksref{CSNSM}\and
E.~Olivieri\thanksref{CSNSM}\and
J.~L.~Ouellet\thanksref{MIT}\and
L.~Pagnanini\thanksref{Milano,INFN-Milano}\and
P.~Pari\thanksref{CEA-IRAMIS}\and
L.~Pattavina\thanksref{LNGS,TUM}\and
B.~Paul\thanksref{CEA-IRFU}\and
M.~Pavan\thanksref{Milano,INFN-Milano}\and
H.~Peng\thanksref{USTC}\and
G.~Pessina\thanksref{INFN-Milano}\and
S.~Pirro\thanksref{LNGS}\and
D.~V.~Poda\thanksref{CSNSM,KINR}\and
O.~G.~Polischuk\thanksref{KINR}\and
E.~Previtali\thanksref{Milano,INFN-Milano}\and
Th.~Redon\thanksref{CSNSM}\and
S.~Rozov\thanksref{JINR}\and 
C.~Rusconi\thanksref{USC}\and
V.~Sanglard\thanksref{IPNL}\and
K.~Sch\"affner\thanksref{LNGS}\and
B.~Schmidt\thanksref{LBNLNSD}\and
Y.~Shen\thanksref{Fudan}\and
V.~N.~Shlegel\thanksref{NIIC}\and
B.~Siebenborn\thanksref{KIT-IK}\and
V.~Singh\thanksref{UCB}\and
S.~Sorbino\thanksref{Sapienza,INFN-Roma}\and
C.~Tomei\thanksref{INFN-Roma}\and
V.~I.~Tretyak\thanksref{KINR}\and 
V.~I.~Umatov\thanksref{ITEP}\and
L.~Vagneron\thanksref{IPNL}\and
M.~Vel\'azquez\thanksref{UGA}\and
M.~Weber\thanksref{KIT-IPE}\and
B.~Welliver\thanksref{LBNLNSD}\and
L.~Winslow\thanksref{MIT}\and
M.~Xue\thanksref{USTC}\and
E.~Yakushev\thanksref{JINR}\and
A.~S.~Zolotarova\thanksref{CSNSM}
}

\thankstext{e1}{e-mail: andrea.giuliani@csnsm.in2p3.fr}

\institute{
IRFU, CEA, Universit\'{e} Paris-Saclay, F-91191 Gif-sur-Yvette, France  \label{CEA-IRFU} \and 
Univ Lyon, Universit\'{e} Lyon 1, CNRS/IN2P3, IP2I-Lyon, F-69622, Villeurbanne, France  \label{IPNL} \and 
National Research Centre Kurchatov Institute, Institute of Theoretical and Experimental Physics, 117218 Moscow, Russia  \label{ITEP} \and 
Dipartimento di Fisica, Sapienza Universit\`a di Roma, P.le Aldo Moro 2, I-00185, Rome, Italy \label{Sapienza} \and 
INFN, Sezione di Roma, P.le Aldo Moro 2, I-00185, Rome, Italy \label{INFN-Roma} \and
Department of Physics, University of California, Berkeley, California 94720, USA \label{UCB} \and
CNRS-N\'eel, 38042 Grenoble Cedex 9, France \label{Neel} \and
Dipartimento di Fisica, Universit\`{a} di Milano-Bicocca, I-20126 Milano, Italy \label{Milano} \and 
INFN, Sezione di Milano-Bicocca, I-20126 Milano, Italy \label{INFN-Milano} \and 
CSNSM, Univ. Paris-Sud, CNRS/IN2P3, Universit\'e Paris-Saclay, 91405 Orsay, France \label{CSNSM} \and
Nikolaev Institute of Inorganic Chemistry, 630090 Novosibirsk, Russia \label{NIIC} \and
LAL, Univ. Paris-Sud, CNRS/IN2P3, Universit\'e Paris-Saclay, 91898 Orsay, France \label{LAL} \and
Laboratory of Nuclear Problems, JINR, 141980 Dubna, Moscow region, Russia \label{JINR} \and
IRAMIS, CEA, Universit\'{e} Paris-Saclay, F-91191 Gif-sur-Yvette, France \label{CEA-IRAMIS} \and
Institute for Nuclear Research, 03028 Kyiv, Ukraine \label{KINR} \and 
Karlsruhe Institute of Technology, Institut f\"{u}r Kernphysik, 76021 Karlsruhe, Germany \label{KIT-IK} \and
Nuclear Science Division, Lawrence Berkeley National Laboratory, Berkeley, California 94720, USA \label{LBNLNSD} \and
DISAT, Universit\`a dell'Insubria, 22100 Como, Italy \label{DISAT} \and
Key Laboratory of Nuclear Physics and Ion-beam Application (MOE), Fudan University, Shanghai 200433, PR China \label{Fudan} \and
Massachusetts Institute of Technology, Cambridge, MA 02139, USA \label{MIT} \and
Karlsruhe Institute of Technology, Institut f\"{u}r Prozessdatenverarbeitung und Elektronik, 76021 Karlsruhe, Germany \label{KIT-IPE} \and 
Physics Division, Lawrence Berkeley National Laboratory, Berkeley, California 94720, USA \label{LBNLPH} \and
INFN, Laboratori Nazionali del Gran Sasso, I-67100 Assergi (AQ), Italy \label{LNGS} \and
Physik Department, Technische Universit\"at M\"unchen, Garching D-85748, Germany \label{TUM} \and
Department of Modern Physics, University of Science and Technology of China, Hefei 230027, PR China \label{USTC} \and
Department of Physics and Astronomy, University of South Carolina, SC 29208, Columbia, USA \label{USC} \and
Universit\'e Grenoble Alpes, CNRS, Grenoble INP, SIMAP, 38402 Saint Martin d'H\'eres, France \label{UGA} 
}

\date{Received: date / Accepted: date}

\maketitle

\begin{abstract}

\cupidmo is a bolometric experiment to search for neutrinoless double-beta decay (\onbb) of \Mo. 
In this article, we detail the \cupidmo detector concept, assembly, installation in the underground laboratory in Modane in 2018, and provide results from the first datasets. The demonstrator consists of an array of 20 scintillating bolometers comprised of $^{100}$Mo-enriched 0.2 kg \lmo crystals. The detectors are complemented by 20 thin cryogenic Ge bolometers acting as light detectors to distinguish $\alpha$ from $\gamma$/$\beta$ events by the detection of both heat and scintillation light signals. We observe good detector uniformity, facilitating the operation of a large detector array as well as excellent energy resolution of 5.3 keV (6.5 keV) FWHM at 2615 keV, in calibration (physics) data. Based on the observed energy resolutions and light yields a separation of $\alpha$ particles at much better than 99.9\% with equally high acceptance for \ga/\be events is expected for events in the region of interest for \Mo \onbb. We present limits on the crystals' radiopurity ($\leq$3~$\mu$Bq/kg of $^{226}$Ra and $\leq$2~$\mu$Bq/kg of $^{232}$Th). Based on these initial results we also discuss a sensitivity study for the science reach of the \cupidmo experiment, in particular, the ability to set the most stringent half-life limit on the \Mo \onbb decay after half a year of livetime. The achieved results show that CUPID-Mo is a successful demonstrator of the technology - developed in the framework of the LUMINEU project - selected for the CUPID experiment, a proposed follow-up of CUORE, the currently running first tonne-scale cryogenic \onbb experiment.

\end{abstract}



\section{Introduction}
\label{sec:intro}

Two-neutrino double-beta decay (\nnbb) is one of the rarest processes in nature.
Initially proposed by Maria Goeppert-Meyer in 1935~\cite{GoeppertMayer:1935}, it has since been observed for 11 nuclei with typical half-lives ranging from 10$^{18}$ to 10$^{24}$\,yr~\cite{Barabash:2015,Barabash:2019}.

Numerous extensions of the Standard Model predict that double-beta decay could occur without neutrino emission (e.g., see \cite{Vergados:2016,DellOro:2016tmg,Bilenky:2015,Deppisch:2012,Rodejohann:2012}). 
This hypothetical transition, called neutrinoless double-beta decay (\onbb), is a lepton-number violating process. Its signature is a peak in the electron sum-energy spectrum at the $Q$-value of the transition (\Qbb). Its observation could help explain the cosmological baryon number excess~\cite{Fukugita:1986}, and would prove that neutrinos are Majorana fermions (i.e., their own antiparticles) \cite{Schechter:1982,DellOro:2016tmg}.

The current leading \onbb decay experiments have a sensitivity on the \onbb half-life of 10$^{25}$--10$^{26}$~yr~\cite{Gando:2016,Albert:2018,Aalseth:2018,Alduino:2018,Agostini:2018}. At present, there is no confirmed observational evidence for \onbb decay, which implies that next-generation experiments have to further increase their discovery potential by at least one order of magnitude.

One of the most promising technologies for \onbb decay searches are cryogenic calorimeters, historically also referred to as bolometers \cite{Fiorini:1984}. These detectors are sensitive to the minute temperature rise induced by energy deposited in a crystal cooled to cryogenic temperatures ($\sim$10\,mK). Key benefits of bolometers are some of the best energy resolutions in the field ($\Delta E(\mathrm{FWHM})/E\sim$ 0.2$\%$), high detection efficiency and the possibility to grow radiopure crystals with a large degree of freedom in the choice of the material.  Dual readout devices i.e. scintillating bolometers further allow for particle identification and thus yield the prospect of studying multiple \onbb decay candidate isotopes in the background free regime \cite{Giuliani:2018,Poda:2017}.

The \cuore (Cryogenic Underground Observatory for Rare Events) experiment~\cite{Alduino:2017pni,Alduino:2018}, currently collecting data at Laboratori Nazionali del Gran Sasso (LNGS, Italy), demonstrates the feasibility of a tonne-scale detector based on this technology. The success of this experiment is the starting point of \cupid (\cuore Upgrade with Particle IDentification), which aims to increase the mass of the \onbb decay isotope via isotopic enrichment while decreasing the background in the region of interest. 

According to the \cuore background model, the dominant background in the \onbb decay region originates from $\alpha$ particles emitted by radioactive contamination of the crystals or nearby materials~\cite{Alduino:2017}. \cupid aims to identify and suppress this background using scintillating crystals coupled to light detectors~\cite{Pirro:2005ar}. A further background suppression can be attained by choosing \onbb decay emitters with a \Qbb well above the 2.6~MeV line of \TL, which is typically the end-point of natural $\gamma$ radioactivity. Bolometers containing isotopes such as \\
\noindent  $^{100}$Mo ($Q_{\beta\beta}$ = 3034.40 $\pm$ 0.17 keV \cite{Rahaman:2008}), \\
\noindent  $^{82}$Se ($Q_{\beta\beta}$ = 2997.9 $\pm$ 0.3~ keV \cite{Lincoln:2012fq}) or \\
\noindent  $^{116}$Cd ($Q_{\beta\beta}$ = 2813.50 $\pm$ 0.13 keV \cite{Barabash:2016}) \\
\noindent  satisfy this condition.

The dual readout concept, where both the heat and light signals are recorded, has been implemented in two medium-scale \cupid demonstrators: \cupido, taking data at LNGS since 2017, and \cupidmo, which started the physics data-taking at the beginning of 2019 in the Laboratoire Souterrain de Modane (LSM, France). Following the \cupid strategy, both experiments make use of enriched crystals (24 \enrZS crystals for \cupido~\cite{Azzolini:2018tum} and 20 \enrLMO crystals for \cupidmo) to search for \onbb decay.

With an exposure of $\sim$10~kg$\times$yr, \cupido proved that dual readout could suppress the dominant $\alpha$ background to a negligible level, obtaining the lowest background level for a bolometric experiment to date~\cite{Azzolini:2019nmi}. Nevertheless, the radiopurity and energy resolution (20.05$\pm$0.34~keV FWHM at \Qbb) of the \cupido crystals \cite{Azzolini:2019nmi,Azzolini:2019tta} do not meet the requirements of \cupid and demand further R$\&$D activity if \enrZS were to be used.

Conversely, the \enrLMO crystals chosen by \cupidmo have demonstrated excellent radiopurity and energy resolution in the tests performed within the LUMINEU (Luminescent Underground Molybdenum Investigation for NEUtrino mass and nature) experiment \cite{Armengaud:2017,Poda:2017a}.

The primary goal of \cupidmo is to demonstrate on a larger scale the reproducibility of detector performance in terms of the high energy resolution and efficient $\alpha$ rejection power combined with high crystal radiopurity. Given the high number of \Mo emitters contained in enriched crystals and the favorable \onbb transition probability for \Mo, \cupidmo also enables a competitive \onbb decay search.

The present work describes the \cupidmo experimental setup, currently operating in the \edw-III~\cite{Armengaud:2017b, Hehn:2016} cryostat at LSM. The detector was constructed in the clean rooms of the Laboratoire de l'Acc\'el\'elarateur Lin\'eaire (LAL) and the Centre de Sciences Nucl\'eaires et de Sciences de la Mati\'ere (CSNSM, Orsay, France) in the fall of 2017 and then moved to LSM and installed in the cryostat in January 2018. The detector was successfully operated through the summer of 2018 (Commissioning I). The fall of 2018 was devoted to cryostat maintenance, after a severe cryogenic failure, and detector upgrades. After optimization of the cryogenic system and detectors over the winter of 2019 (Commissioning II), the experiment has been collecting data in a stable configuration since the end of March 2019 (Physics run). In this paper we present the \cupidmo detector concept and construction (Sec.~\ref{sec:setup}), the operation and initial performance of the first Physics run dataset (Sec. \ref{sec:performance}), and the prospects of the experiment in \onbb decay search (Sec. \ref{sec:outlook}).

\section{Experimental setup}
\label{sec:setup}

\begin{figure}[htbp]
  \centering
  \includegraphics[width=0.49\textwidth]{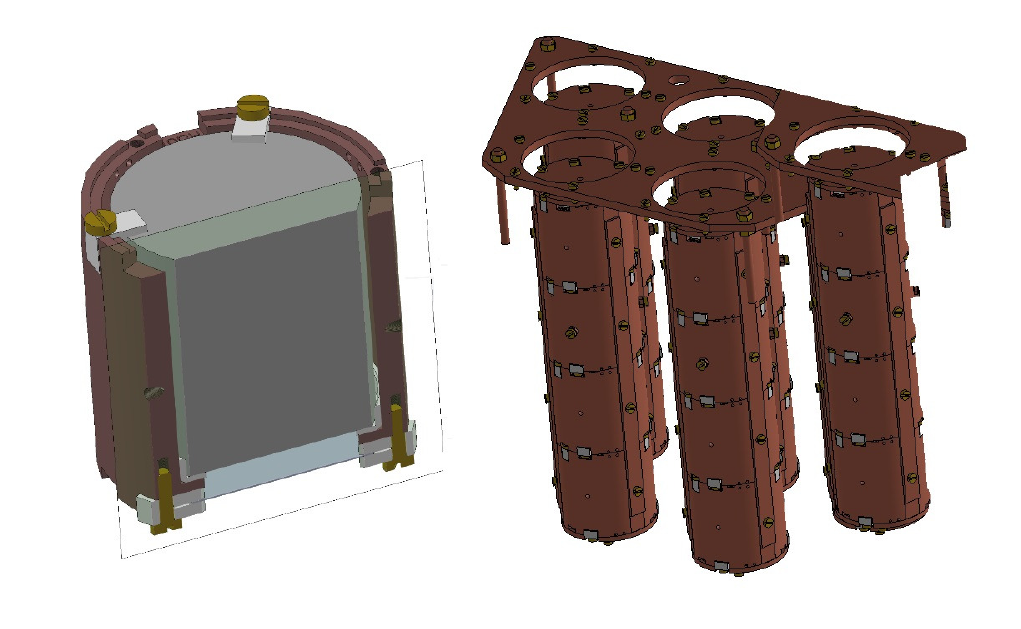}
  \caption{Rendering of a \cupidmo single detector module (left) designed to hold a \enrLMO scintillating element. A module is comprised of a crystal of size $\oslash$44$\times$45~mm and a Ge wafer of $\oslash$44$\times$0.175~mm. The full 20-detector bolometric array is arranged in five suspended towers containing four detector modules each (right).}
  \label{fig:design}
\end{figure}

\cupidmo consists of an array of 20 scintillating bolometer modules arranged in five towers, each with four modules, as shown in Fig.~\ref{fig:design}.
Each module contains one \enrLMO crystal and one germanium wafer
assembled inside a single-piece copper housing,  instrumented with Neutron Transmutation Doped (NTD) Ge thermistors.
All the materials used for the towers' construction
were carefully selected, and additionally cleaned as needed to minimize radioactive contamination.
The detector construction, transportation, and assembly into the underground cryogenic facility were performed in a clean environment. The key ingredients of the detector, its assembly, and the cryogenic apparatus are detailed below.

\subsection{\enrLMO crystals}

\cupidmo operates the four existing \enrLMO crystals previously used in LUMINEU~\cite{Armengaud:2017,Poda:2017a}.
An additional sixteen new \enrLMO crystals were fabricated with the identical procedure as that employed by LUMINEU~\cite{Berge:2014,Armengaud:2017,Grigorieva:2017}.
All crystals have a cylindrical shape with $\sim44$~mm diameter and $\sim45$~mm height, and a mass of $\sim0.2$~kg. The crystals were produced at the Nikolaev Institute of Inorganic Chemistry (NIIC, Novosibirsk, Russia) as follows:
\begin{itemize}
\item purification of the $\sim97\%$ enriched molybdenum~\cite{Berge:2014}, previously used in the NEMO-3 experiment \cite{Arnold:2015};
\item selection of lithium carbonate with low U/Th and $^{40}$K content~\cite{Armengaud:2017} and purified \Mo oxide~\cite{Grigorieva:2017};
\item crystal growth via a double crystallization process using the low-thermal-gradient Czochralski technique~\cite{Grigorieva:2017,Armengaud:2017};
\item slicing of the scintillation elements, and treatment of their surfaces with radio-pure SiO powder.
\end{itemize}
The total mass of the 20 \enrLMO crystals used in \cupidmo is 4.158~kg, corresponding to 2.264~kg of \Mo.

\subsection{Ge slabs}

The high-purity Ge wafers (Umicore Electro-Optical Material, Geel, Belgium), used as absorbers for the scintillation light, have a diameter of 44.5~mm and a $175~\upmu$m thickness. A $\sim$70~nm SiO coating was evaporated on both sides of the Ge wafer to make them opaque, thus increasing the light collection by $\sim$35\% \cite{Mancuso:2014}. A small part of the wafer surface was left uncoated to ease the gluing of a temperature sensor.

\subsection{Sensors}
\label{sec:sensors}

\cupidmo employs NTD Ge thermistors~\cite{Haller:1994} as thermal sensors. These thermistors were provided by the Lawrence Berkeley National Laboratory (LBNL, Berkeley, USA) and come from a single production batch. The \ntd{}s used for the \lmo (LMO) bolometers are $3.0\times3.0\times1.0$\,mm$^{3}$ in dimension, and have a temperature-dependent resistance given by $R = R_0 \cdot e^{(T_0/T)^{0.5}}$\ where the average values for the parameters are $T_0=3.8$~K and  $R_0=1.5~\Omega$. Given the lower heat capacity of the Ge absorbers for the bolometric light detector (LD), we opted to better match and reduce the  heat capacity of their sensors by dicing the \ntd{}s into multiple pieces.

In Commissioning I, we produced three sensors with $3.0\times0.8\times0.4$\,mm$^{3}$ dimensions
from the slicing of a single \ntd in two directions.
The LDs with these sensors showed an unexpectedly high noise with a strong $1/f$[Hz] component reaching frequencies up to hundreds of Hz. For Commissioning II and beyond, we replaced all but two sensors with new ones with $3.0\times0.8\times1.0$\,mm$^{3}$ dimensions, avoiding the horizontal cut of the original \ntd{}s.  

In addition to the thermistor, each \enrLMO crystal is instrumented with a silicon-based resistive chip \cite{Andreotti:2012} operated as a heater. This heater allows us to periodically inject a constant power and generate a pulse of constant energy. The resulting reference pulses can be used in the offline analysis to monitor and correct for a change of the signal gain due to temperature drifts of the bolometer \cite{Alessandrello:1998}.

\subsection{Sensor coupling}
\label{sec:couplings}

\begin{figure}
\centering
\includegraphics[width=0.49\textwidth]{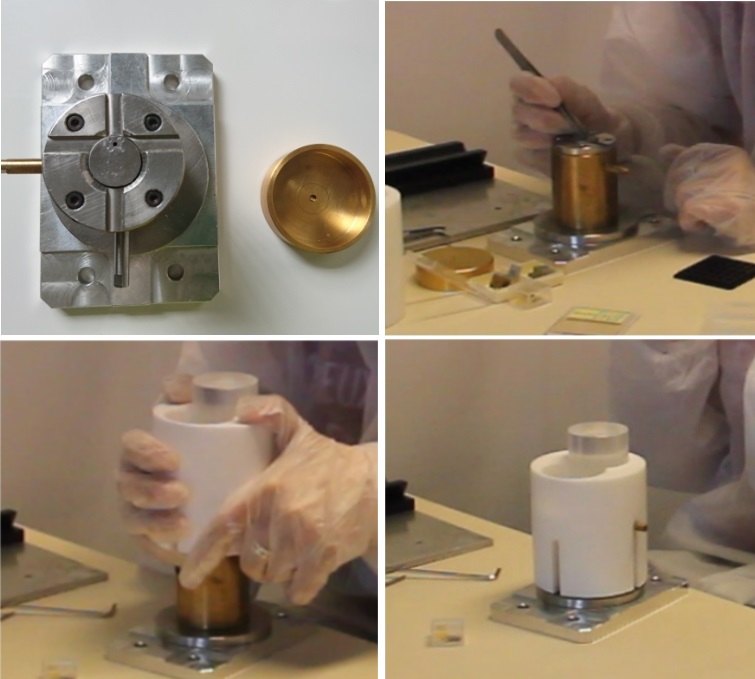}
\caption{The tool for gluing the sensor (top left) together with the cap to provide a 50 $\mu$m gap. The process of gluing is shown in the three other photos: \ntd placement (top right) and placement of the crystal (bottom panels).}
\label{fig:gluing_tool}
\end{figure}

The \ntd{}s were glued on the \enrLMO crystals using the dedicated tool shown in Fig.~\ref{fig:gluing_tool}, similar to the one used by \cupido \cite{Azzolini:2018tum}. The glue is a two-component epoxy resin (Araldite{\textregistered} Rapid) well-tested for cryogenic applications and demonstrated to have acceptable radiopurity \cite{Alduino:2017}. The gluing tool features a part for holding the \ntd, and it can be moved along the vertical axis and fixed at any level. The performance of the bolometer is strongly dependent upon the quality of gluing, and we obtain optimal results when separate glue spots connect the \ntd to the crystals. This helps in compensating the different thermal contractions of the involved materials.
 To maintain separate glue spots during and after the epoxy curing, the \ntd is kept 50~$\upmu$m from the \enrLMO crystal, which is positioned on the top surface of the gluing tool. \ntd{}s of five crystals (LMO-1--4,15), used in the LUMINEU experiment and/or the \cupidmo single tower test, were glued with six spots, while nine spots were applied for the remaining crystals. The heaters were glued with a single glue spot using a 50~$\mu$m Mylar mask to provide a gap between the crystal surface and the chip. 

The gluing of \ntd{}s to the Ge wafer was also performed with the two-component epoxy resin described above. However, instead of the six or nine spot matrix, we applied a uniform veil of glue. This choice was motivated by the small size of the sensor and the less pronounced effect of thermal contraction expected for the Ge-glue-Ge interface.  We used the manipulator of an ultrasonic bonding machine to provide a controlled force to attach the \ntd to the Ge wafer surface and provide better reproducibility.

\subsection{Detector structure}
\label{sec:DetectorModule}

The \cupidmo single module and tower structure were designed by the Service de Physique de l'Etat Condens\'e (SPEC) at CEA (Commissariat \`a l'\'Energie Atomique et aux \'energies alternatives, Gif-sur-Yvette, France) according to the following requirements:

\begin{itemize}
\item the single module structure should be compact but permit the housing of four scintillating crystals in a single tower, taking into account the restricted space in the experimental set-up;
\item the towers should be suspended by dedicated springs to mitigate the vibrational noise of the set-up~\cite{Armengaud:2017};
\item the design should allow a simple installation inside the cryostat (see Fig. \ref{fig:design} and Sec. \ref{sec:EDELWEISS}).
\end{itemize}

\begin{figure}
\centering
\includegraphics[width=0.45\textwidth]{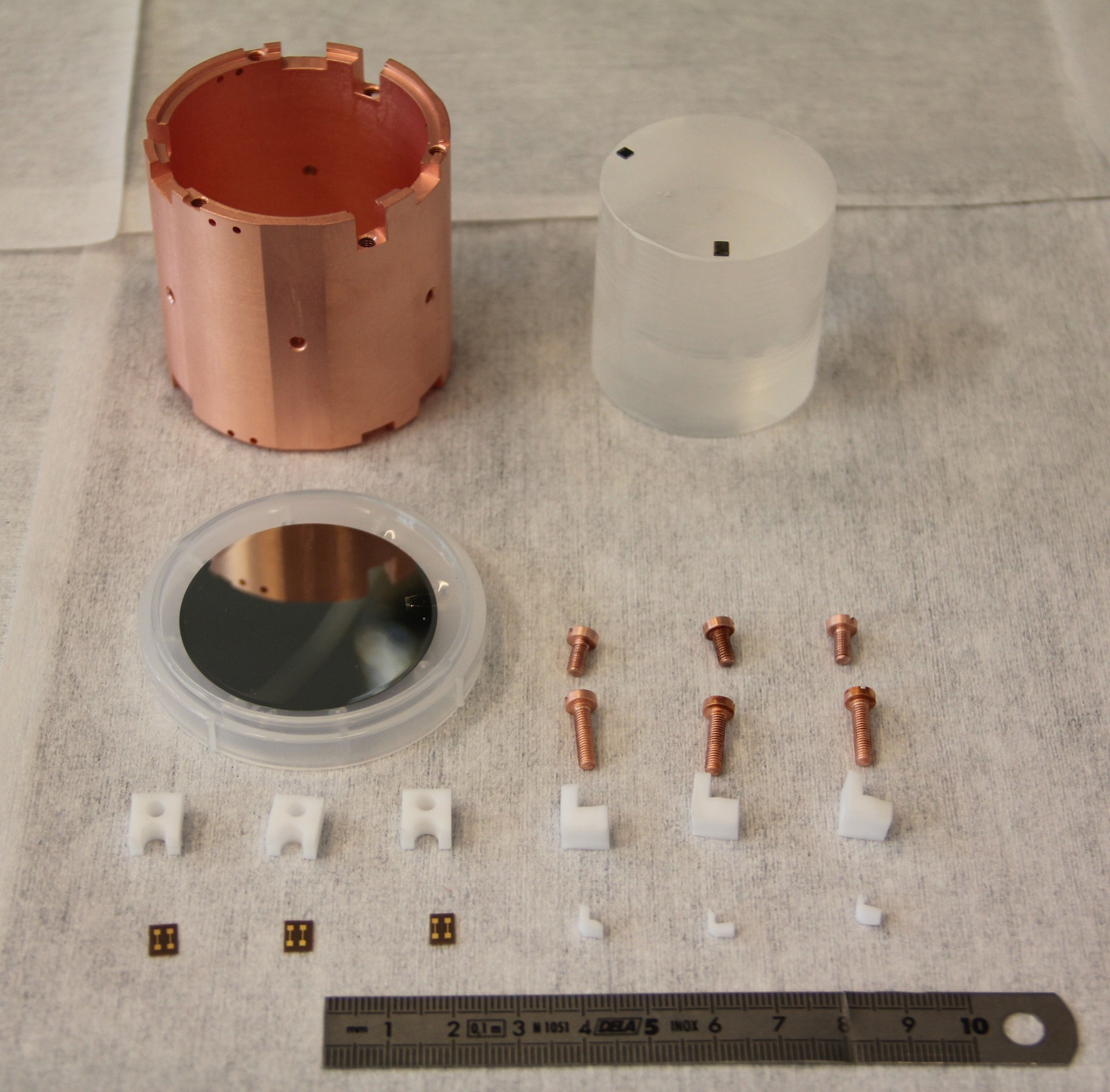}
\caption{All components used to assemble a \cupidmo single detector module: a copper holder, a \enrLMO crystal with glued \ntd and heater, a Ge LD with \ntd, the copper screws, the PTFE spacers and fixing elements, the Kapton foil with golden pads. Note that the reflecting film is not shown here.}
\label{fig:components}
\end{figure}

The  mechanical workshop of LAL (Orsay, France) fabricated the detector support structure. Each detector module (see Fig. \ref{fig:components}) is a single-piece holder, made of highly radiopure NOSV{\texttrademark}  copper from Aurubis (Hamburg, Germany). It contains both a \enrLMO scintillation element and a Ge wafer.
The bolometers are kept in place by small Polytetrafluoroethylene (PTFE) holders which decouple them from the thermal bath. The \enrLMO crystals are supported with three PTFE elements on the top and bottom, while the LDs are clamped with three PTFE pieces. In Commissioning I, we did not install any reflecting foil around the crystals, because previous measurements demonstrated efficient particle identification performance despite a factor of 2 lower light collection efficiency~\cite{Poda:2017}. Commissioning I was characterized by sub-optimal LD performance (see Sec.~\ref{sec:sensors}), hence we decided to surround the crystals' lateral side with reflecting foil (3M Vikuiti{\texttrademark}) in addition to the replacement of the LDs' \ntd{}s.

\subsection{Detector assembly}
\label{sec:Assembly}

\begin{figure}
\centering
\includegraphics[width=0.45\textwidth]{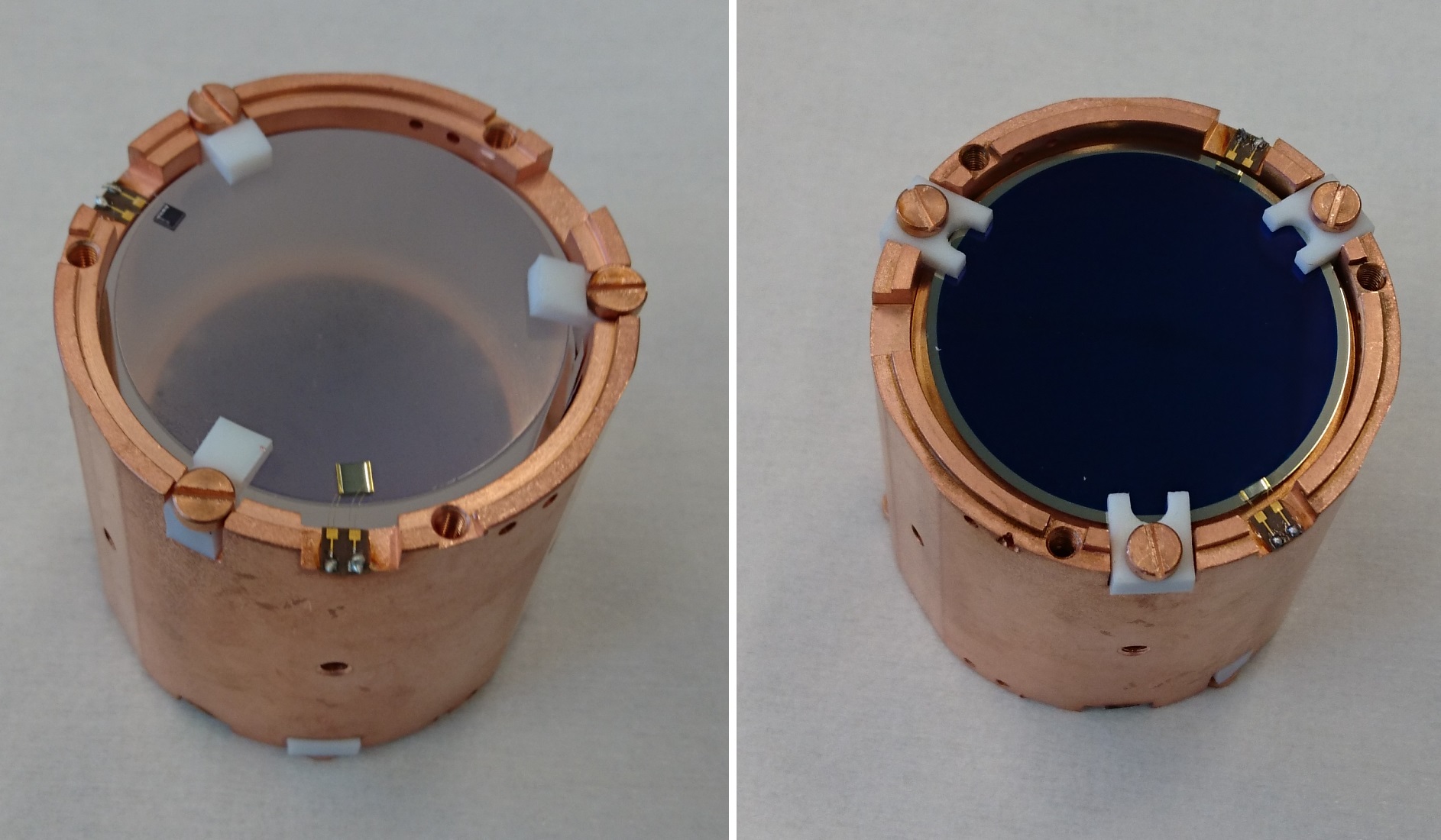}
\caption{An assembled \cupidmo module. On the left: view from the top, the semi-transparent crystal surface where the \ntd and heater are located. On the right: view from the bottom, germanium LD with SiO coating (dark blue internal circle); the 2 mm on the edge of the wafer (17\% of the area) are uncoated.}
\label{fig:module}
\end{figure}

\begin{figure}
\centering
\includegraphics[width=0.45\textwidth]{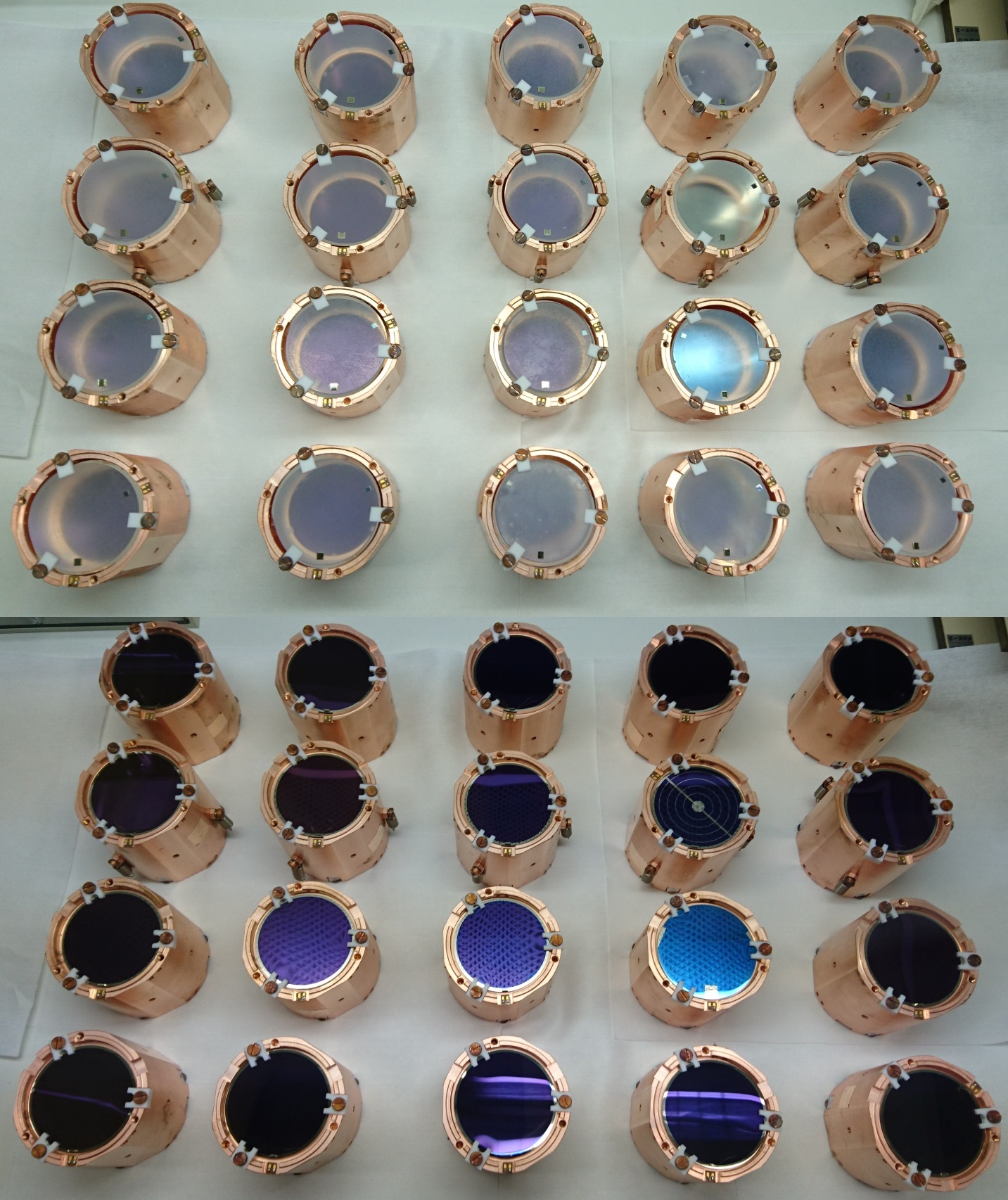}
\caption{Twenty CUPID-Mo detector modules before their arrangement in five towers; the devices are shown from the \enrLMO crystal side (top) and from the Ge LD side (bottom).}
\label{fig:ALLmodules}
\end{figure}

\begin{figure}[hbt]
\centering
\includegraphics[width=0.45\textwidth]{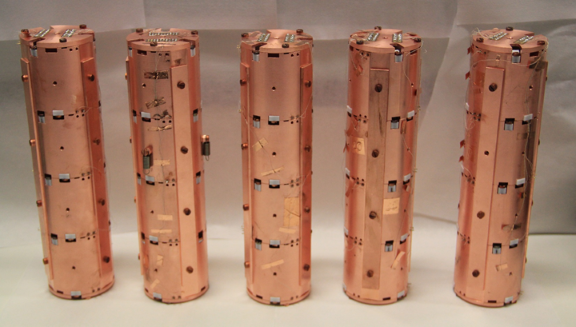}
\caption{The five assembled towers before installation in the cryogenic facility.}
\label{fig:5towers}
\end{figure}

We performed all activities related to the detectors' assembly in a cleanroom environment. All the used detector components were carefully cleaned before assembly to minimize the risk of surface re-contamination. 
In particular, copper elements were etched with citric acid, and PTFE elements cleaned with ethanol in an ultrasonic bath. The tower assembly was performed in a class 10 cleanroom at LAL.
The \lmo crystal is fixed inside its copper housing with PTFE elements on top and bottom as well as surrounded by a reflecting foil. An assembled single module (top and bottom view), as well as all the \cupidmo detectors, are shown in  Fig.~\ref{fig:module} and Fig. \ref{fig:ALLmodules}, respectively.
In total, five towers were assembled with four detectors each (Fig.~\ref{fig:5towers}).
Crystals on the lower three floors are each viewed by one LD from the top and one LD from the bottom. The top floor crystals are each viewed by one LD from the bottom.

\subsection{Wiring}
\label{sec:Wiring}

A dedicated wiring scheme was designed and implemented for the \cupidmo experiment as the existing \edw-III readout could not accommodate the additional 20 dual-readout modules required for \cupidmo.

We bonded gold wires from the \ntd{}s to flat Kapton pads with gold contacts to provide the electrical readout connection as well as the weak thermal link to the heat bath.
Silk-covered constantan twisted wires were soldered on the other side and run up each tower to a larger Kapton pad with gold contacts glued at the top of the tower.

On this pad the constantan wires and copper wires (connection to the cold electronics)  were soldered\footnote{Later, high purity Ge counting (HPGe) revealed that the solder is not Pb-free (the activity of $^{210}$Pb is (170$\pm$15) Bq/kg); the total mass of the material on the towers is (1.2$\pm$0.1) g.}. This connection provides a link to Si-JFET (junction gate field-effect transistor) based pre-amplifiers at 100\,K through the copper plate inside the \edw cryostat (see Sec. \ref{sec:EDELWEISS}).

\subsection{Low background cryogenic facility}
\label{sec:EDELWEISS}

\begin{figure}[b]
\centering
\includegraphics[width=0.45\textwidth]{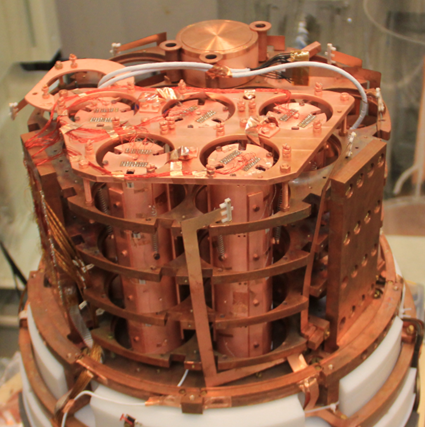}
\caption{The \cupidmo scintillating bolometer array installed inside the \edw set-up. The remainder of the experimental volume is occupied by eleven Ge-based bolometers of the \edw direct dark matter search experiment and one scintillating bolometer for CUPID R\&D.}
\label{fig:EDW}
\end{figure}

The \cupidmo detector array is installed (see Fig. \ref{fig:EDW}) in the \edw-III cryogenic set-up \cite{Armengaud:2017b,Armengaud:2017}, located in LSM. This site is among the deepest underground laboratories in the world; the 1700 m (4800 m water equivalent) rock overburden, provided by the Frejus mountain, reduces the cosmic muon flux to 5 muons/m$^2$/day \cite{Schmidt:2013}. 

\begin{figure}[b]
\centering
\includegraphics[width=0.45\textwidth]{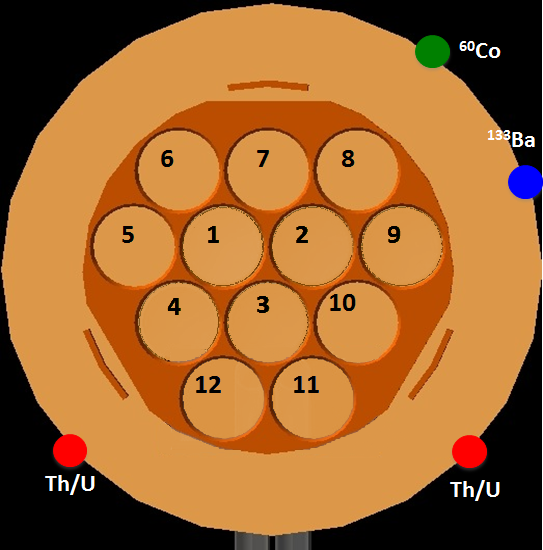}
\caption{A schematic top view on detectors plate inside the \edw set-up showing twelve slots for tower installation inside copper plates. The position and types of calibration sources are also indicated.}
\label{fig:EDWslots}
\end{figure}

The \edw cryostat is a custom dilution refrigerator with a reversed geometry \cite{Armengaud:2017b}, developed by Institut N\'eel (Grenoble, France). During a cryogenic run, this set-up requires periodic refilling of the liquid helium (LHe) bath every 10 days. The consumption of LHe is minimized by the use of a cold vapor reliquefaction system based on three Gifford-MacMahon cryocoolers. The cryocoolers are responsible for most of the vibrational noise in the set-up, thus necessitating the use of the suspension to achieve high-performance operation of the scintillating bolometers \cite{Armengaud:2017}. The passive shielding of the set-up against environmental radiation consists of lead (20 cm thickness) and polyethylene (55~cm thickness). The inner part of the lead shield is made of 2\,cm thick low $^{210}$Pb radioactivity ($<0.12$~Bq/kg) lead recovered from sunken Roman-era galleys (hereafter called "Roman lead"). An additional internal Roman lead (14~cm) and polyethylene (10~cm) shield at the 1K-plate is used to protect the detectors from radioactivity from the cryostat components. A muon veto system is surrounding the whole cryostat providing 98\% geometrical coverage. The muon veto is constructed from 46 individual plastic scintillator modules with a total surface of 100 m$^2$ and provides a detection efficiency of 97.7\% for muons passing through a central sphere with 1~m radius \cite{Schmidt:2013}. The set-up is located inside a class 10000 cleanroom with a depleted radon air supply ($\sim$30 mBq/m$^3$ of $^{222}$Rn). 

The experimental volume of \edw-III contains four floors (detector plates) with twelve slots each (see Fig.~\ref{fig:EDWslots}). The \cupidmo towers were inserted through the slots T3, T4, T10, T11, and T12 (see Table~\ref{tab:Towers}) and mechanically decoupled from the \edw-III detector plate with three metal springs for each tower\footnote{The activity of $^{228}$Th and $^{226}$Ra and $^{60}$Co in the springs is $\sim$0.01~Bq/kg, while the $^{40}$K content is $\sim$4~Bq/kg; the total mass of the material is around 8 g.}. The remaining experimental space is partially occupied (tower slots T2, T5, T7, and T8) by 11 Ge bolometers for the \edw dark matter search program \cite{Arnaud:2018} and a cadmium tungstate based scintillating bolometer for CUPID R\&D \cite{Helis:2019}.

\begin{table}[hbt]
\centering
\caption{\cupidmo tower compositions and their corresponding slots inside the \edw set-up at LSM. }
\footnotesize
\begin{center}
\begin{tabular}{c|c|c|c|c|c}
\hline\noalign{\smallskip}
Tower & \multicolumn{5}{c}{Detectors' IDs and towers' slots}     \\
\cline{2-6}
composition & T3 & T4 & T10 & T11 & T12 \\
\hline
LMO & 5 & 9 & 17 & 2 & 13  \\
LD & ~ & ~ & ~ & ~ & ~  \\
\hline
LMO & 6 & 10 & 16 & 3 & 18  \\
LD & ~ & ~ & ~ & ~ & ~  \\
\hline
LMO & 7 & 11 & 14 & 4 & 19  \\
LD & ~ & ~ & ~ & ~ & ~  \\
\hline
LMO & 8 & 12 & 1 & 15 & 20  \\
LD & ~ & ~ & ~ & ~ & ~  \\
\hline
\end{tabular}
 \label{tab:Towers}
\end{center}
\end{table} 

We added two mixed U/Th sources made of thorite mineral to the \edw-III automatic source deployment system \cite{Armengaud:2017b} (see Fig.~\ref{fig:EDWslots}).  These sources complement the already available $\gamma$-calibration sources of $^{133}$Ba ($\sim$1~kBq) and $^{60}$Co  ($\sim$100~kBq)  for periodic calibration of the \cupidmo detectors. The activities of the sources are $\sim$50 Bq of $^{232}$Th, $\sim$100 Bq of $^{238}$U, and few Bq of $^{235}$U. The $^{133}$Ba source emits $\gamma$s with energies up to 0.4\,MeV and was only used during the commissioning stage. The high-activity $^{60}$Co $\gamma$ source is used to eliminate space charges in the dual readout heat-ionisation Ge bolometers of \edw-III \cite{Armengaud:2017b} and also to calibrate the \cupidmo Ge LDs via source-induced X-ray fluorescence \cite{Poda:2017d,Berge:2018} (see Sec.~\ref{sec:LDperformance}). 
The $^{60}$Co source is used mainly during and just after each LHe refill (every 10 days) while a regular $\sim$2-days-long Th/U calibration is scheduled for each period between subsequent LHe refills.

The detector readout in the \edw-III setup is based on AC-biased cold electronics \cite{Armengaud:2017b}, which restricts the use of high-resistivity thermistors to at most a few~M$\Omega$ resistance at a given bias current (working point) \cite{Armengaud:2017}. Custom made room-temperature electronics modules called bolometer boxes (BBs) are mounted just outside of the cryostat to ensure short cables to limit noise pick-up. These BBs contain the electronics for the cold Si-JFET pre-amplifiers' biasing, Digital to Analog Converters (DACs) for the detectors' biasing, post-amplification, anti-aliasing filter, and ADCs to record the \cupidmo \ntd{}s \cite{Armengaud:2017b}. All LMOs and five LDs are operated with BBs containing 16-bit ADCs, while the signal digitization for the remaining LDs is done with 14-bit ADCs. The pulser system, used to inject a constant Joule power through the heaters, is based on a 4-channel pulse generator with a typical injection periodicity of a few minutes. The data acquisition system \cite{Armengaud:2017b,Armengaud:2017} can record both online triggered and stream data; the triggered data is used only for monitoring purposes.

\section{\cupidmo{} operation and performance}
\label{sec:performance}

\subsection{\cupidmo{} detector operation}

Of the 20 LMO and LD pairs, only a single LD was lost due to a hardware issue, resulting in 39 out of 40 active channels. Additionally, 18 out of 20 heaters are available to inject pulses. The optimal working point of the LMO detectors was chosen to maximize the signal amplitude. LDs, instrumented with smaller, more resistive, sensors operate in an over-biased regime to obtain an \ntd resistance of $\sim$1~M$\Omega$ mitigating the impact of AC biasing (see details in \cite{Armengaud:2017}). The modulation frequency of 500 Hz was chosen to reduce the pick-up of cryocooler-induced high-frequency noise\footnote{A better noise environment was observed using the \edw wiring \cite{Armengaud:2017b} in LUMINEU allowing for a 1 kHz modulation frequency \cite{Armengaud:2017,Poda:2017a}.}. 
The nominal base temperature of the empty \edw cryostat is 11.5\,mK. In the present, densely populated, cryogenic setup, an additional heat load increases this base temperature to $\sim$20\,mK and we could stably operate at  20.7\,mK with a few $\mu$W of regulation power.
This temperature  is considerably higher than the operating temperature in the LUMINEU predecessor \cite{Armengaud:2017} and it is expected to have an adverse effect on the detector performance.

Nevertheless, the following analysis of a $\sim$ 2 week period with 11.1 days of physics data, 2.2 days of mixed Th/U source calibration, and 1.6 days of \Co irradiation provides a robust confirmation of the bolometric performance achieved within LUMINEU \cite{Armengaud:2017, Poda:2017}. The data were acquired between March 24th 2019 and April 6th 2019 and correspond to a physics exposure of 0.1 \ky \enrLMO. 
This early data is comparable with the prior exposure presented in \cite{Poda:2017}, and emphasizes the reproducibility of \lmo detectors  using a total of 20 detectors.

\subsection{Data processing}
\label{sec:processing}

Two independent analysis frameworks, both exploiting the optimum filter technique \cite{Gatti:1986}, are used for the data processing: one, called DIANA \cite{Alduino:2016, Azzolini2018b}, is adapted from the CUORE \cite{Alduino:2018} and CUPID-0 \cite{Azzolini2018b} experiments and the other was developed at CSNSM \cite{Mancuso:2016} and used for the analysis of the LUMINEU data \cite{Armengaud:2017}. The CSNSM code has been developed (using the MATLAB MULTI Integrated Development Environment) specifically for the analysis of scintillating bolometer data. It is more nimble and readily adapted to different experimental setups. In contrast, DIANA is a much broader framework, including analysis packages for larger detector arrays (in particular allowing for the analysis of coincident events). It is object-oriented C++ code with a  PostgreSQL \cite{Stonebraker:1986} database interface to track detector and electronics settings. The use of DIANA allows for comparison between different \cupid project demonstrator experiments with effectively the same analysis tools, and DIANA is expected to be used as the primary package for \cupidmo in the future.
Therefore, all results presented below are based on the use of DIANA, while it is noted that very similar results were obtained with the CSNSM code, providing a cross-check of the DIANA processing.

\subsection{Performance of bolometric Ge light detectors}
\label{sec:LDperformance}

Characteristic pulse shape parameters such as the rise- and decay-times, defined as 10\% to 90\% of the rising edge and 90\% to 30\% of the trailing edge of the LD pulse shape have been investigated (see Table \ref{tab:LDperformance}). We estimate typical (median) rise- and decay-times of 4.2\,ms and 9.2\,ms respectively from an averaged pulse, triggered and aligned on events re\-corded in an associated LMO crystal.

Averaging of  pulses was necessary since \lmo has a moderate Relative Light Yield (\rly) which does not exceed 1 keV/MeV relative to the heat signal (see \cite{Armengaud:2017,Poda:2017a} and Sec. \ref{sec:LMOperformance}), and estimates from individual light pulses are subject to bias from noise fluctuations. We note that in particular for the rise-time both the 500 Hz sampling and the alignment of the average pulse become limiting factors for a more precise estimate. At a previous surface test at CSNSM with a similar temperature, working point, and a 10\,kHz sampling rate, a factor of 3 faster rise-time (0.96\,ms) was observed in LD 4.

\begin{table}
\centering
\caption{Performance of Ge light detectors of the CUPID-Mo experiment. Detectors marked with an asterisk (*) suffer from additional uncertainty due to prominent sinusoidal noise. 
The quoted parameters are the \ntd resistance at the working point (R$_{Work}$), the rise time ($\tau_R$), the decay time ($\tau_D$), the voltage sensitivity ($A_{Signal}$), and the baseline noise resolution (FWHM$_{Noise}$). For the definition of the listed variables see text.}
\footnotesize
\begin{center}
\begin{tabular}{c|c|c|c|c|c}
\hline\noalign{\smallskip}
LD & R$_{Work}$  & $\tau_R$ & $\tau_D$ & $A_{Signal}$ & FWHM$_{Noise}$    \\
ID & (M$\Omega$) & (ms) & (ms) & ($\mu$V/keV) & (eV) \\
\hline
1 & 1.18 & 5.0 & 8.3 & 0.5 & 183  \\ 
2 & 0.87 & 3.6 & 9.2 & 1.2 & 92  \\ 
3 & 0.37 & 3.3 & 5.6 & 0.5 & 1225  \\ 
4 & 0.92 & 3.2 & 8.5 & 1.7 & 175  \\ 
5 & 0.75 & 3.5 & 9.9 & 1.1 & 66  \\ 
6 & 0.70 & 3.8 & 9.5 & 1.0 & 175  \\  
7 & -- & -- & -- & -- & --  \\ 
8 & 0.50 & 3.8 & 3.6* & 0.7 & 368  \\ 
9 & 0.95 & 3.6 & 4.7* & 1.0 & 354  \\ 
10 & 0.91 & 3.5 & 5.2* & 1.1 & 323  \\
11 & 1.83 & 4.5 & 4.2 & 2.6 & 146  \\ 
12 & 0.86 & 5.2 & 18.0 & 1.0 & 69  \\ 
13 & 0.70 & 9.7 & 11.5 & 1.3 & 112  \\ 
14 & 0.90 & 9.0 & 10.2 & 1.2 & 122  \\ 
15 & 0.77 & 4.3 & 14.1 & 1.2 & 86  \\ 
16 & 0.46 & 4.2 & 9.4 & 0.8 & 219  \\ 
17 & 0.73 & 4.2 & 7.2 & 1.3 & 129  \\ 
18 & 0.66 & 4.1 & 9.3 & 1.2 & 121  \\ 
19 & 0.76 & 4.3 & 9.0 & 1.2 & 202  \\ 
20 & 1.14 & 4.9 & 23.5 & 0.6 & 137  \\ 
\hline
Median & 0.77 & 4.2 & 9.2 & 1.1 & 146  \\ 
\hline
\end{tabular}
 \label{tab:LDperformance}
\end{center}
\end{table} 

\begin{figure}[htbp]
\centering
\includegraphics[width=0.49\textwidth]{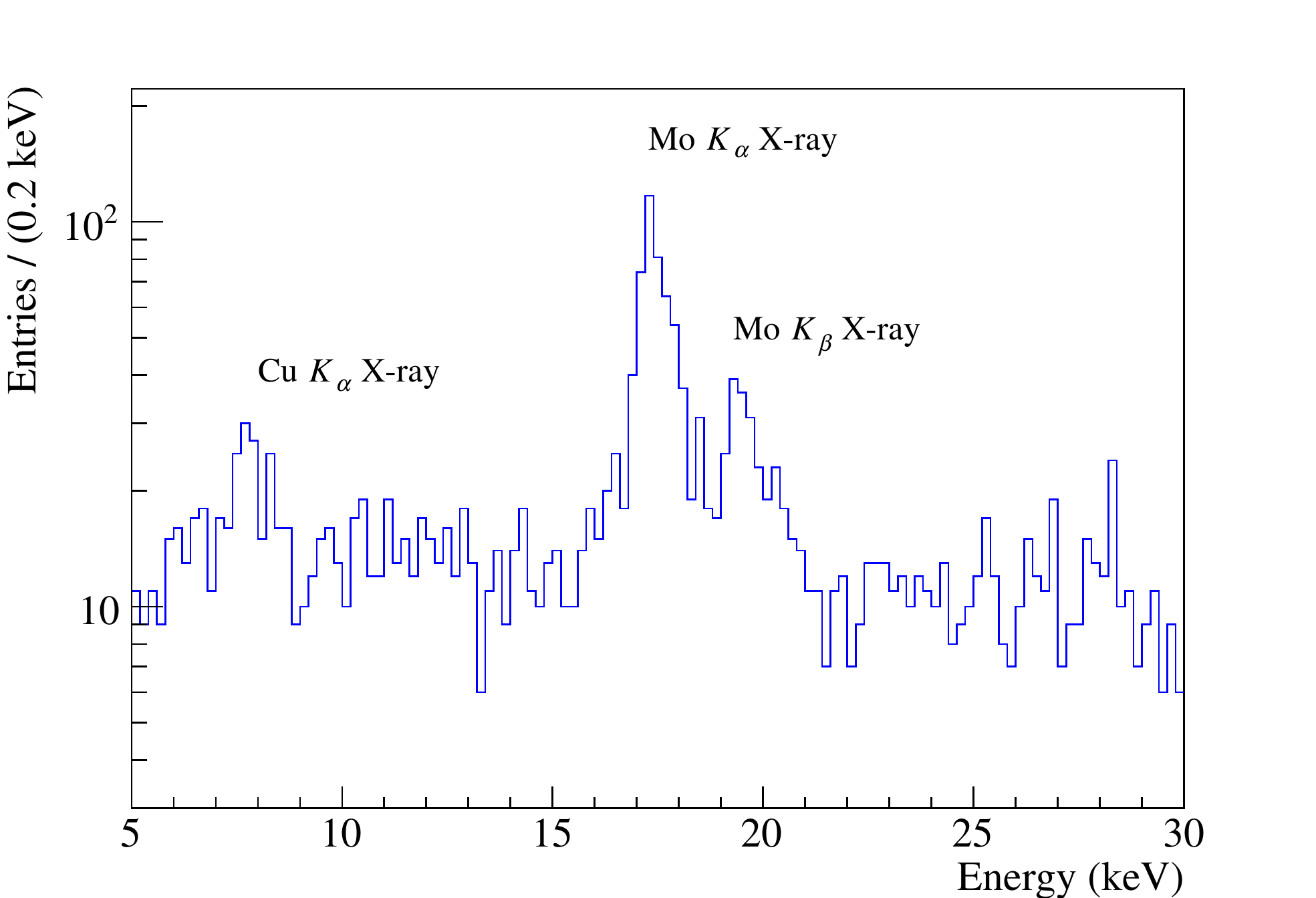}
\caption{Energy spectrum of CUPID-Mo light detector LD2 after a 33 h \Co irradiation in the EDELWEISS-III set-up. The Mo K$_{\alpha}$ \xr{}s are used for the LD calibration. }
\label{fig:CoLdCalibration}
\end{figure}

To estimate the performance of the Ge LDs, we perform an in situ calibration. We employ the X-ray fluorescence of Mo or Cu that is generated when the crystals and setup are exposed to a higher intensity $\gamma$ source \cite{Poda:2017d,Berge:2018}. For Mo we expect characteristic peaks from the K$_{\alpha1}$\ (17.48 keV, intensity $I=100\%$), K$_{\alpha2}$\ (17.37 keV, $I=52\%$), and K$_{\beta1}$\ (19.61 keV, $I=15\%$) lines \cite{Thompson:2009}. 
The Cu \xr{}s can give additional peaks from K$_{\alpha1}$\ (8.05 keV, $I=100\%$), K$_{\alpha2}$\ (8.03 keV, $I=51\%$), and K$_{\beta1}$\ (8.91 keV, $I=17\%$).  
Fig. \ref{fig:CoLdCalibration} shows a typical X-ray spectrum obtained during the \Co source irradiation. The prominent features are a sum K$_{\alpha}$ peak from Cu and both a sum K$_{\alpha}$ and a distinct K$_{\beta}$ peak from Mo. The intensity of the Cu \xr{}s is much lower than those associated with Mo, as the Cu is only facing the LDs on the side. Also, the statistics in the Cu K$_{\alpha}$ peak are very low for detectors far from the \Co source, and we chose to omit this peak from the LD calibration. 

With a stable operating temperature of 20.7\,mK and a strong \ntd polarization for the Ge LDs, negligible nonlinearity is expected. We use a Gaussian fit to the most intense peak, the Mo K$_{\alpha}$ \xr{}s, and perform a first-order polynomial calibration with zero intercepts. 

The 1.4\,g Ge LDs are instrumented with small-size \ntd{}s that achieve a typical sensitivity of 1.1 $\mu$V/keV with an RMS of $\sim$ 40\% (see Table \ref{tab:LDperformance}). Uncertainties in the individual sensitivity estimates are dominated by the gain in the analog chain with typical uncertainties of order 10\% for several of the operational amplifiers in the amplification chain. 

The LDs sensitivity is limited by a comparatively high regulation temperature of the detector plate and the strong \ntd polarization\footnote{A factor of two higher signal amplitude in LD 4, operated at 17 mK, was observed in a surface test at CSNSM (see Fig. 23 in \cite{CUPIDInterestGroup:2019inu}).}.

We estimate the baseline resolution for all detectors from a set of forced random trigger events injected every 101\,s. We exclude one detector instrumented with a different \ntd sensor (used in Commissioning I), see Table \ref{tab:LDperformance} and runs with atypical noise performance, resulting in 183/209 (LD-bolo\-meter, run) pairs. 
The median of these estimates yields a typical baseline resolution of 148\,eV FWHM in agreement with the channel based estimate in Table \ref{tab:LDperformance}. 
We see good reproducibility with individual channel estimates ranging from 66\,eV up to 368\,eV. 

\begin{figure}[htbp]
\centering
\includegraphics[width=0.49\textwidth]{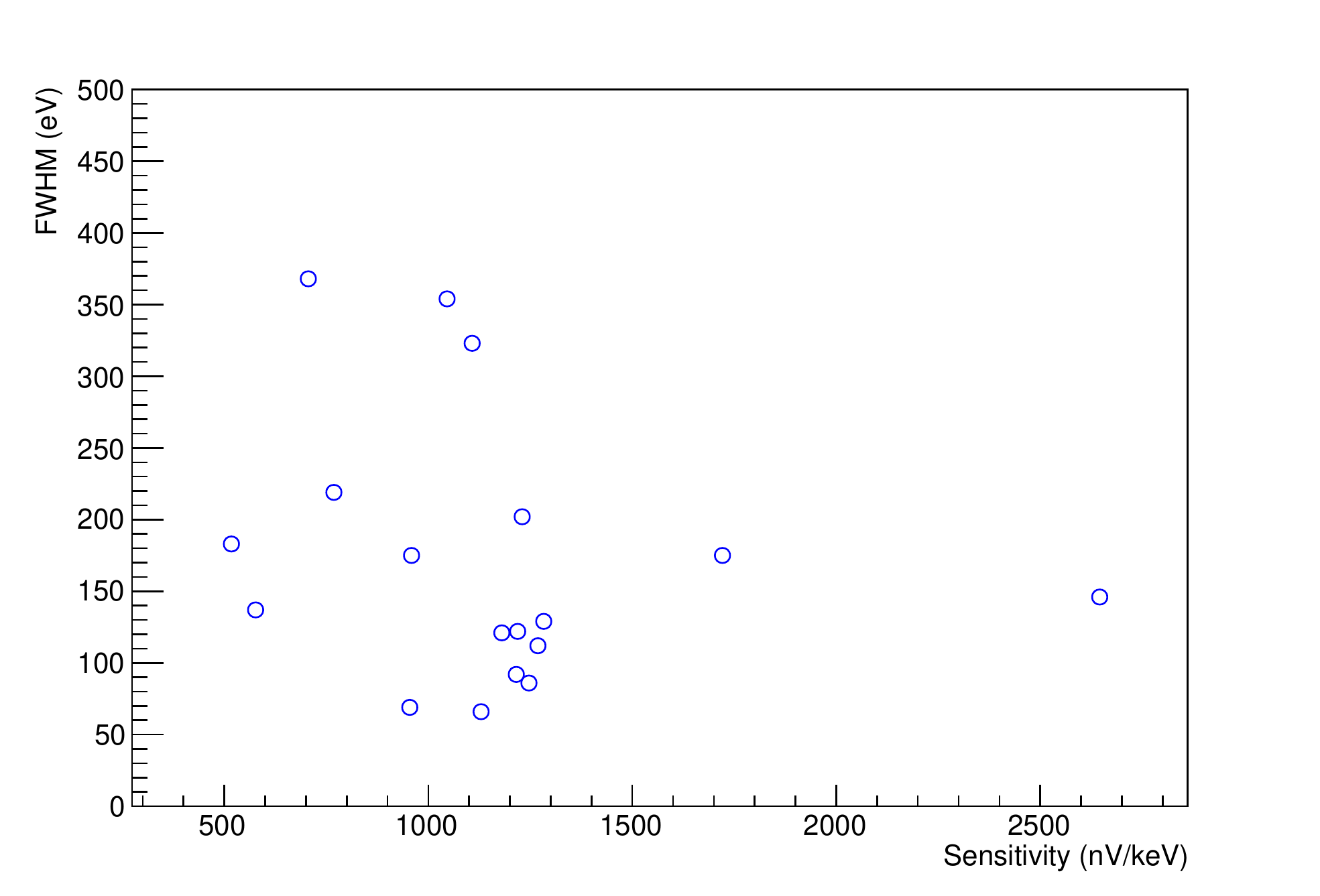}
\caption{The baseline resolution versus sensitivity for 18/20 Ge LDs operated in the CUPID-Mo experiment at 20.7 mK. One detector is discarded due to having a different \ntd, while the second is not operational (see text).}
\label{fig:LDSensitivityFWHM}
\end{figure}

A resulting scatter plot of the correlation between the sensitivity and the achieved baseline resolution is shown in Fig. \ref{fig:LDSensitivityFWHM}. 
We note that the spread in detector performance is only slightly higher than for the \ntd{}s on the \lmo crystals (see Sec.~\ref{sec:LMOperformance}). We want to emphasize the uniformity that lends itself to applications in larger cryogenic detector arrays.

For reference, we list the performance characteristics on an individual LD basis in Table \ref{tab:LDperformance}. 
The reported performance in terms of the baseline resolution exceeds the requirements to achieve a better than 99.9\% rejection of \al events at 99.9\% acceptance of $\gamma$/$\beta$s as is discussed in detail in Sec. \ref{tab:LMOperformance}. 
Several improvements can be pursued for the full-size CUPID experiment.
DC-biased electronics, higher sampling rate, and the implementation of  additional analysis and de-noising techniques can improve the quoted performance. 
Futhermore, lower noise NTD Ge sensors and a lower operational temperature resulting in a higher detector sensitivity can also yield a significantly better LD performance, as demonstrated with a 20 eV FHWM baseline resolution  in \cite{Barucci:2019}.

\subsection{Performance of \enrLMO{} bolometers}
\label{sec:LMOperformance}

The time constants of LMO bolometers are much longer than those of LDs. We obtain median values of 24\,ms for the rise-time and 299\,ms for the decay-time with a significant spread of 208\,ms in the decay-times and a smaller spread of 8\,ms in the rise-time (see Table \ref{tab:LMOperformance}). These values are consistent with previously reported values \cite{Armengaud:2017}, and in the typical range for macroscopic cryogenic bolometers operated in the tens of mK range.

We calibrate with a mixed Th/U source, with a most prominent peak at 2615\,keV (\TL) and negligible gamma continuum, see Fig. \ref{fig:SumCalibLMO}. This is the closest observable \ga-line, $\sim$ 415 keV lower than the $Q_{\beta\beta}$-value of \Mo.

The calibration data were acquired over a short period (2.2 days), resulting in limited statistics of the detected $\gamma$ peaks. We neglect nonlinearities in the detector response and fit using zero and the 2615\,keV \TL line. Additionally, we use this peak to correct for changes in thermal gain due to slow temperature drifts in the experimental setup. The resulting correction is a linear scaling factor obtained from the optimum filter (OF) amplitude versus baseline dependence in the calibration data and is applied to both calibration and background data. 

The detector sensitivity at the 20.7\,mK operation temperature has a median value of 17\,nV/keV with an RMS of about 30\% (see Table \ref{tab:LMOperformance}). For unknown reasons, the detector LMO 2 shows very low sensitivity in comparison to the results of the CUPID-Mo Commissioning I (6 nV/keV at 20.5 mK) and LUMINEU (47 nV/keV at 17 mK \cite{Poda:2017d}). As in the case of LDs, larger sensitivity of LMO bolometers is expected at colder temperatures (e.g., compare results given in \cite{Armengaud:2017}).

\begin{figure}[htbp]
\centering
\includegraphics[width=0.49\textwidth]{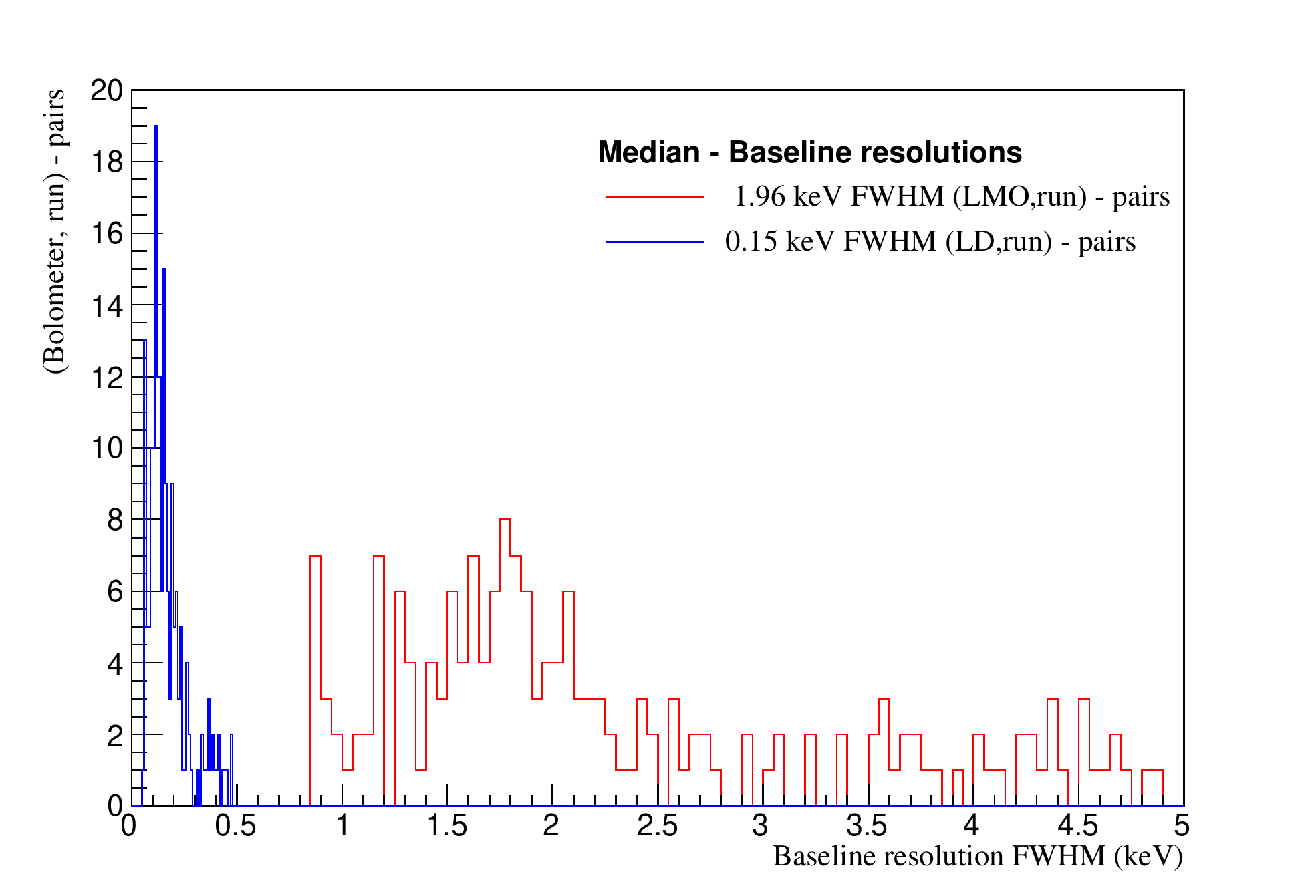}
\caption{Baseline resolution (FWHM) distributions for 183 (LD, run) and 189 (LMO, run) pairs. LMO 2 and LD 3 are rejected, as are periods with atypical noise.}
\label{fig:FWHMDistribution}
\end{figure}

\begin{figure}[htbp]
\centering
\includegraphics[width=0.49\textwidth]{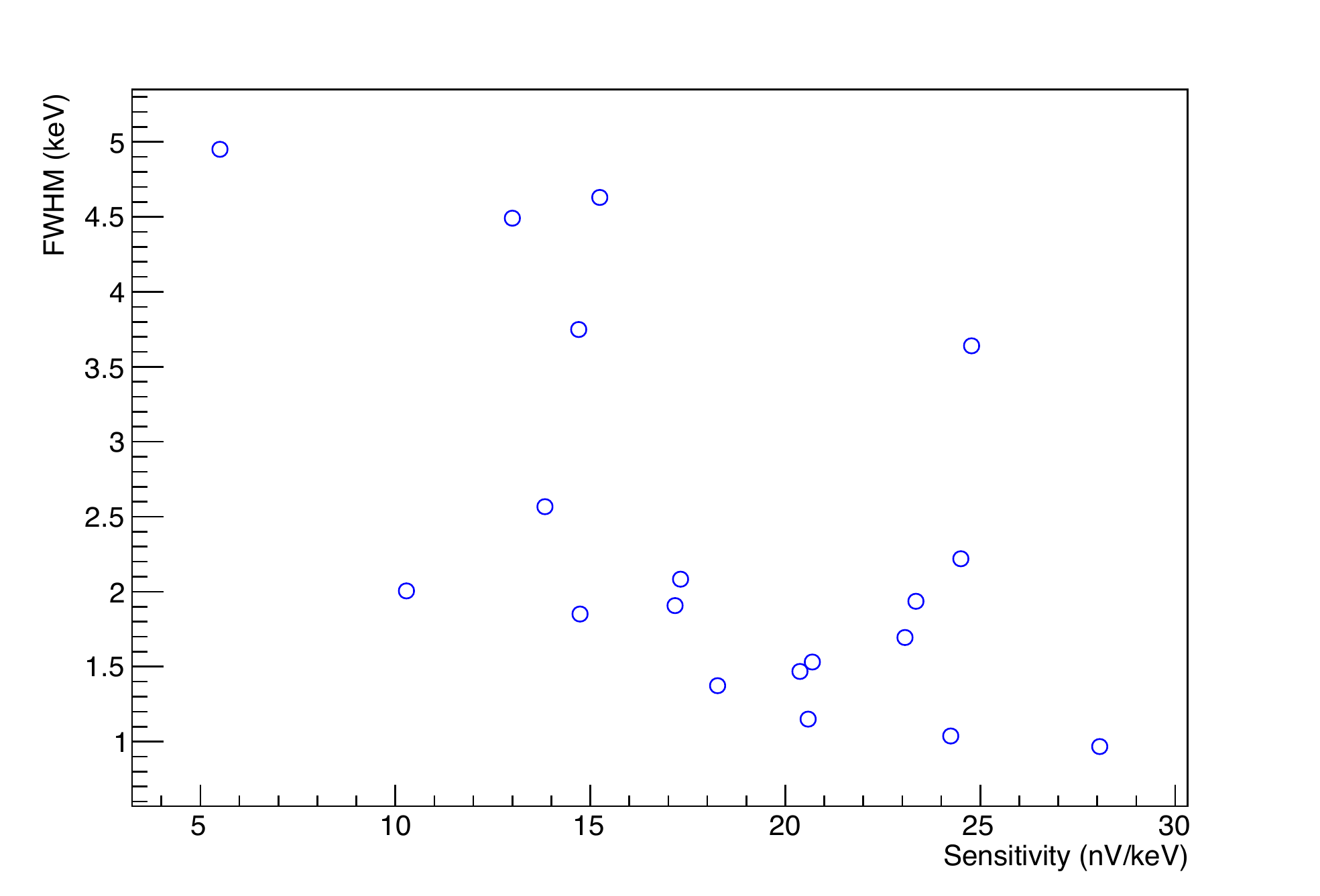}
\caption{Sensitivity versus baseline resolution (FWHM) for the 19 LMO detectors considered in this analysis. One detector (LMO 2) is omitted due to abnormal performance; see text.}
\label{fig:SensitivityLMOFWHM}
\end{figure}

\begin{figure}[tbp]
\centering
\includegraphics[width=0.52\textwidth]{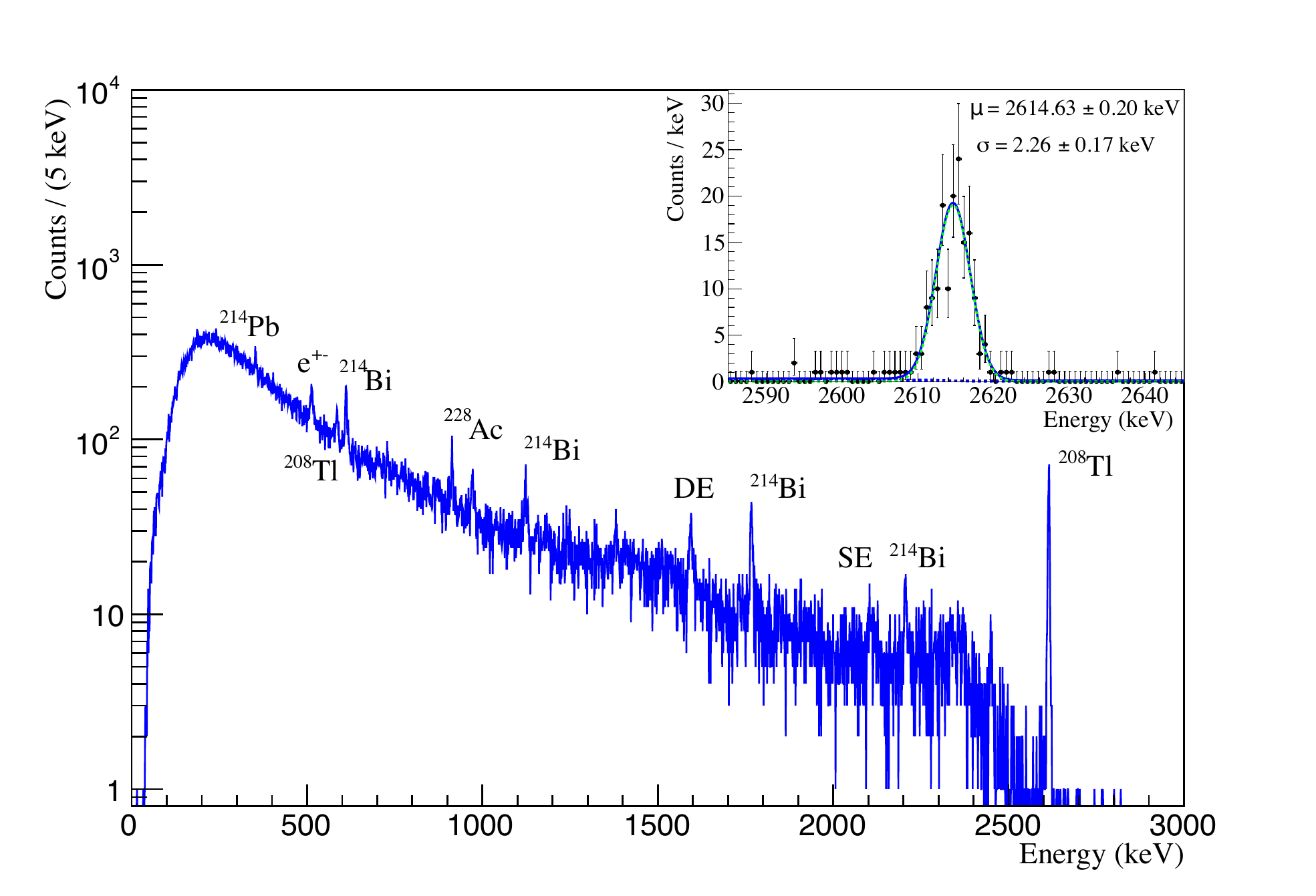}
\caption{Summed calibration spectrum for 19/20 \lmo bolometers. 
All the major peaks have been labeled. The inset shows a fit of the $^{208}$Tl $\gamma$ peak at 2614.5 keV.}
\label{fig:SumCalibLMO}
\end{figure}

The same method utilized for the investigation of LDs' baseline resolution (see Sec.\ref{sec:LDperformance}) is also applied for \enrLMO bolometers. We obtained characteristic (median) values of 1.96\,keV FWHM for the baseline resolution with the spread of the distribution given in Fig. \ref{fig:FWHMDistribution} and individual detector based resolutions presented in Table \ref{tab:LMOperformance}. The baseline noise versus sensitivity data is also illustrated in Fig.~\ref{fig:SensitivityLMOFWHM}.

\newpage
\clearpage

\begin{landscape}

\begin{table}
\centering
\caption{Performance of $^{100}$Mo-enriched \lmo bolometers of the \cupidmo experiment operated at 20.7\,mK in the \edw set-up at LSM (France). This table contains the following information: the crystal size and mass, the \ntd resistance at the working point (R$_{Work}$), the rise-time ($\tau_R$), the decay-time ($\tau_D$), the voltage sensitivity ($A_{Signal}$), the baseline noise resolution (FWHM$_{Noise}$), the scintillation light yield (\rly) measured by top LD ($RLY_{Top}$) and bottom LD ($RLY_{Bottom}$), and the light yield quenching for alpha particles ($QF_{\alpha}$). The omission of a measured parameter due to lack of statistics or insufficient performance / non-operational light detector is indicated by ``--''. The median value for $RLY_{Bottom}$ is given for scintillators coupled to two LDs ($^a$) and for single LD ($^b$); see text.}
\footnotesize
\begin{center}
\begin{tabular}{c|c|c|c|c|c|c|c|c|c|c}
\hline\noalign{\smallskip}
enrLMO & Size & Mass & R$_{Work}$  & $\tau_R$ & $\tau_D$ & $A_{Signal}$ & FWHM$_{Noise}$  &  $RLY_{Top}$ & $RLY_{Bottom}$ & $QF_{\alpha}$ \\
ID & (mm) & (g) & (M$\Omega$) & (ms) & (ms) & (nV/keV) & (keV) & (keV/MeV) & (keV/MeV) &  (\%) \\
\hline
1 & $\oslash43.6\times40.0$ & 185.86 & 1.37 & 31 & 476 & 10 & 2.01 &  0.67 & 0.66 & 19.5 \\
2 & $\oslash43.6\times44.2$ & 203.72 & 1.04 & 48 & 1093 & 1.2 & 30.6 & n.a. & 0.96 & 19.7 \\
3 & $\oslash43.9\times45.6$ & 212.61 & 0.75 & 33 & 302 & 14 & 2.56 &  0.83 & --  & 20.8 \\
4 & $\oslash43.9\times44.5$ & 206.68 & 1.07 & 29 & 264 & 25 & 2.23  & 0.61 & 0.56 & 21.4 \\
5 & $\oslash43.8\times45.0$ & 211.10 & 1.77 & 19 & 584 & 21 & 1.52 &  n.a. & 0.80  & 18.6\\
6 & $\oslash43.8\times45.0$ & 209.19 & 1.69 & 23 & 384 & 25 & 3.62 &  0.65 & 0.64  & 19.1 \\
7 & $\oslash43.8\times45.0$ & 210.45 & 0.85 & 24 & 357 & 21 & 1.15 &  0.77 & --  & 20.9\\
8 & $\oslash43.8\times45.0$ & 209.71 & 1.53 & 29 & 406 & 24 & 1.04 & -- & 0.60 & 19.3 \\
9 & $\oslash43.8\times45.3$ & 209.30 & 0.69 & 28 & 464 & 13 & 4.48  &  n.a. & 0.90 & 19.6 \\
10 & $\oslash43.8\times45.3$ & 208.90 & 1.94 & 32 & 341 & 28 & 0.98 &  0.75 & 0.66  & 21.8 \\
11 & $\oslash43.8\times45.3$ & 208.18 & 2.99 & 18 & 173 & 23 & 1.70 &  0.74 & 0.61  & 17.7\\
12 & $\oslash43.8\times45.4$ & 209.98 & 3.76 & 21 & 213 & 15 & 1.85 &  0.75 & 0.65 & 18.9 \\
13 & $\oslash43.8\times45.5$ & 210.35 & 1.37 & 25 & 445 & 15 & 4.62 &  n.a. & 0.91 & 19.2 \\
14 & $\oslash43.8\times44.5$ & 205.76 & 1.16 & 15 & 95 & 15 & 3.77 &  0.63 & 0.55 & 20.6 \\
15 & $\oslash43.7\times45.2$ & 209.50 & 1.24 & 28 & 195 & 6 & 4.98 &  0.65 & 0.58 & 18.8 \\
16 & $\oslash43.8\times45.0$ & 208.96 & 1.38 & 21 & 228 & 20 & 1.47 &  0.67 & 0.58 & 21.0 \\
17 & $\oslash43.8\times45.3$ & 209.20 & 1.51 & 24 & 292 & 23 & 1.94 &  n.a. & 0.80 & 20.1 \\
18 & $\oslash43.8\times45.3$ & 210.09 & 2.22 & 18 & 192 & 17 & 2.09 &  0.77 & 0.64 &20.0 \\
19 & $\oslash43.8\times45.3$ & 209.16 & 1.36 & 20 & 173 & 18 & 1.38 & 0.76 & 0.68 & 20.4 \\
20 & $\oslash43.8\times45.3$ & 209.70 & 2.48 & 14 & 97 & 17 & 1.89  &  0.67 & 0.66 & 19.2 \\
\hline
Median & $\oslash43.8\times45.3$ & 209.25 & 1.37 & 24 & 297 & 17 & 1.97  &  0.74 & 0.64$^a$ & 19.7 \\
~ & ~ & ~ & ~ & ~ & ~ & ~ & ~  &  ~ & 0.90$^b$ & ~ \\
\hline
\end{tabular}
 \label{tab:LMOperformance}
\end{center}
\end{table} 

\end{landscape}

\newpage

For further analysis, we utilize a preliminary set of analysis cuts. First, periods of atypical noise and temperature spikes of the cryostat are rejected, removing $\sim$11\% of the data from the commissioning period.
A large part of the loss of livetime is caused by a suboptimal setting of the cryostat suspension, and improved stability has been observed in more recent data\footnote{About $\sim$~95\% of the data are kept after the data quality selection since April 2019.}.
We exclude pile-up events with another trigger in a $(-1, +2)$\,s window, require a baseline slope consistent with the typical behavior of the channel, 
and require both the rise-time as well as the optimum filter peak position to be within 5 median-absolute deviations (MAD) of the mean range as defined by the overall distribution of these values.
We further select \ga/\be events by requiring events to have a \rly (see Sec.~\ref{sec:LYperformance}) within 4$\sigma$ of the mean amplitude incident in a LD associated with a LMO bolometer.

The resulting calibration data are presented as a summed spectrum in Fig.~\ref{fig:SumCalibLMO}. 
The 2615 keV \TL resolution is 5.3\,keV FWHM estimated with an unbinned extended maximum likelihood (UELM) fit shown in the inset. The fit model includes a Gaussian function and two components, a smeared step function for multi-Compton events and a locally flat background.
We note a potential bias on the resolution since we perform the thermal gain stabilization on this gamma peak and are in a low statistics limit. A toy Monte-Carlo (MC) with a typical value of 20 counts per detector resulted in an estimated bias (underestimate of the \TL peak width) of 0.3\,keV.

In addition to the good energy resolution, we highlight the linearity and uniformity of the data. The maximum residual between observed peak position and expected peak position in the summed calibration spectrum was 3\,keV for the 1120\,keV line from \BI.
Similarly, we observe an excess width for all \ga peaks of at most 5\,keV due to not yet accounted for individual detector non-linearities.

\subsection{Performance of light-vs-heat dual readout}
\label{sec:LYperformance}
We estimate the \rly from events in the 2--3\,MeV region, close to the $Q$-value for \onbb of \Mo. We create a distribution of light/heat energies and fit a Gaussian to this distribution to obtain the \rly $\mu_{\gamma}$ for \ga/\be events.
We obtain 31 individual (LMO, LD) pairs comprised of 15 LMOs in the line of sight of two LDs (minus a failed, and an under-performing LD) and 5 LMOs with a direct line of sight to a single LD (see Sec.~\ref{sec:Assembly}).

\begin{figure}[htbp]
\centering
\includegraphics[width=0.52\textwidth]{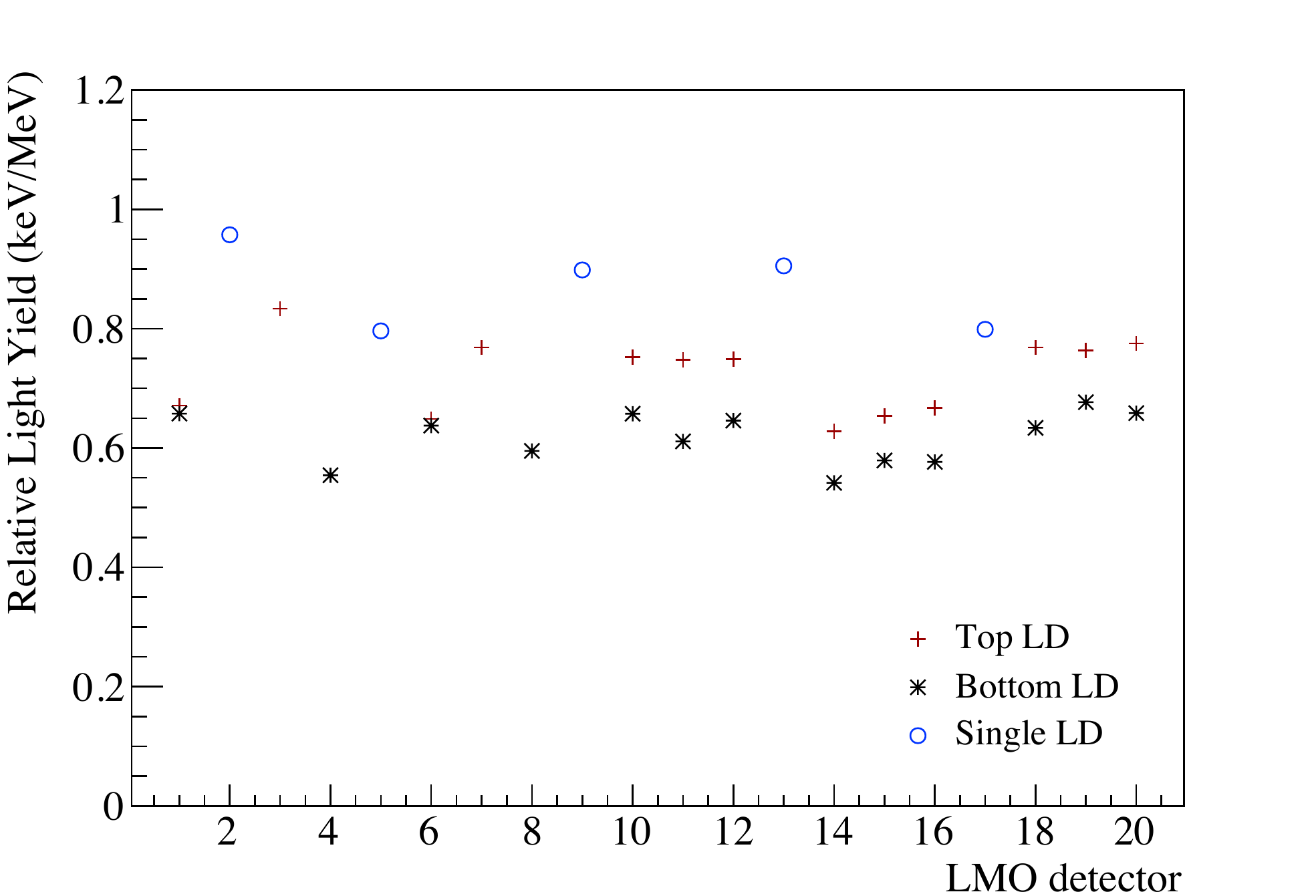}
\caption{Relative Light Yields for all of the \enrLMO crystals. The differences in \rly trace the expected light collection efficiencies due to the design of the towers, see Sec. \ref{sec:DetectorModule} and \ref{sec:Assembly}.}
\label{fig:LY_LMO}
\end{figure}

The resulting \rly{}s (in keV/MeV) are listed in Table \ref{tab:LMOperformance} and illustrated in Fig.\ref{fig:LY_LMO}. 
The obtained \rly includes both the effect of scintillation light production in the crystal as well as light propagation to the Ge absorber, and a pattern that is dominated by the latter effect emerges.  
 \enrLMO crystals on the top of the towers with a reflective copper cap at the upper side and a single LD at the bottom, observe the highest RLYs with a median value of 0.90\,keV/MeV. In addition a $\sim$0.1\,keV/MeV difference is observed in light collection between the top (0.74\,keV/MeV) and the bottom (0.64\,keV/MeV) LDs. 
This effect is a result of a protrusion that is part of the crystal support, which acts as an aperture for downward going light. The obtained results are consistent with previous observations \cite{Armengaud:2017,Poda:2017a}. The summed light collected from two adjacent LDs is the closest estimate we have for ideal light collection. It is as high as 1.44\,keV/MeV with a median value of 1.35\,keV/MeV. 
The uncertainty for individual \rly estimates has been quantified from the spread in \rly estimates of three distinct \Co plus \TL datasets. We observed a $\sim$4\% spread around the mean (RMS), with a maximum deviation of 16\% for a single detector.

For this analysis, we opt to use the LD in the same detector module just below the crystal by default. In cases where the lower LD is unavailable or performs significantly worse (LMO 1, 3, 6 and 7) we switched to associating the upper LD to this crystal (see also Tables \ref{tab:LDperformance} and \ref{tab:LMOperformance}).

\begin{figure}[htbp]
\centering
\includegraphics[width=0.52\textwidth]{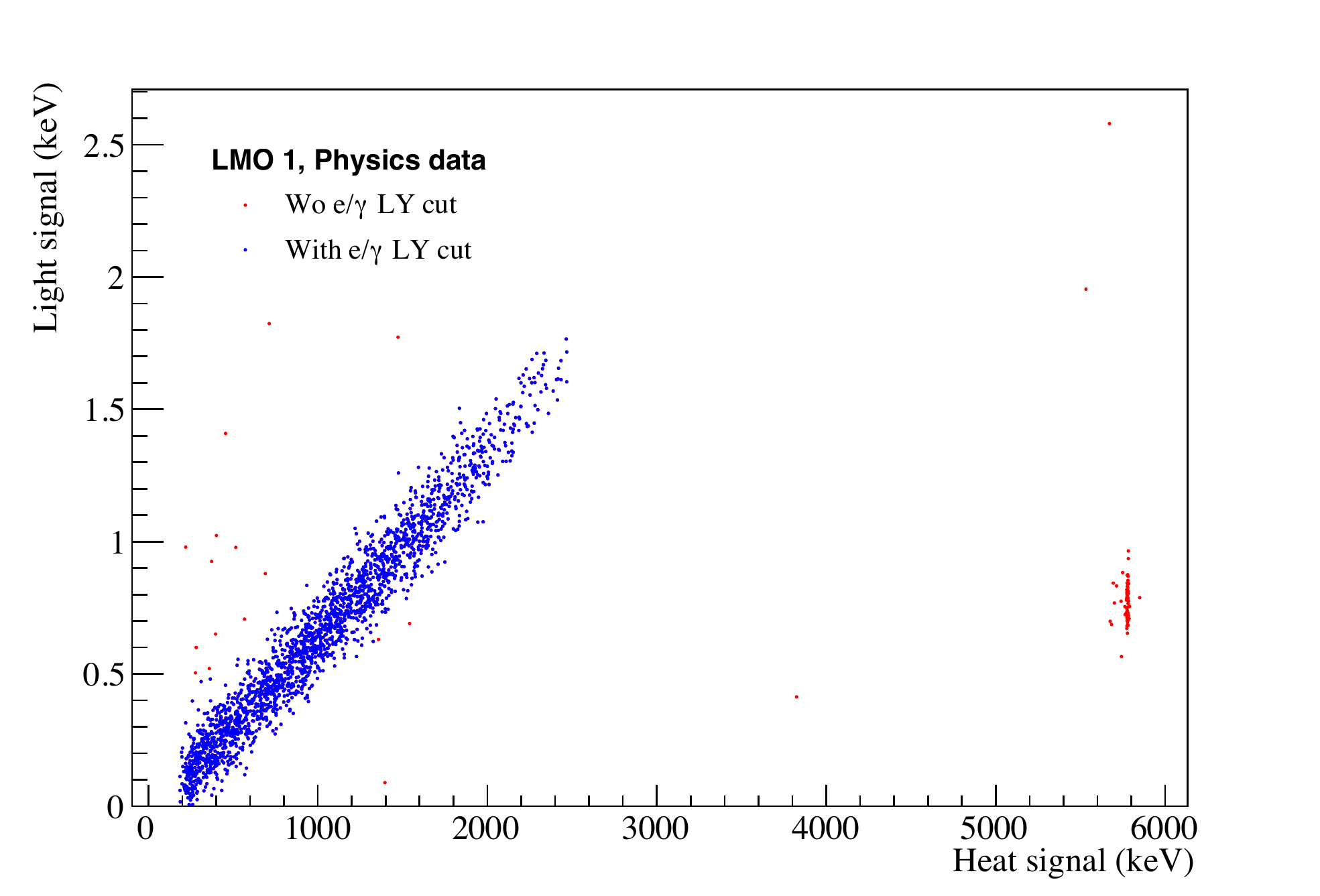}
\caption{Light yield versus heat signal scatter-plot using 11 days of physics data from LMO 1. The presented detector has the highest \PO contamination to illustrate best the distributions of \al and \ga events and the scintillation light quenching for \al events. }
\label{fig:ExampleLightHeatBg}
\end{figure}

Taking into account the measured \rly (Table \ref{tab:LMOperformance}) and the LD performance (Table \ref{tab:LDperformance}), all detectors achieve better than 99.9\% discrimination of \al events (see sec. \ref{sec:AlphaDiscrimination}) with a typical example of the discrimination power given in Fig. \ref{fig:ExampleLightHeatBg}. The preliminary \ga/\be selection by \rly (blue) defined before eliminates a significant population of \al events with $\sim$20\% of the \rly of \ga/\be events and a few remaining events at higher light yield than expected (red).  
This particular crystal is characterized by the highest contamination level of \PO with $\sim$ 0.5 mBq/kg and hence best exemplifies the alpha discrimination power achieved for a scintillating bolometer with typical performance values of 0.67 keV/MeV \rly and 0.18 keV FWHM$_{Noise}$ of a coupled LD. 
We observe that the \PO $\alpha$ events misreconstructed at $\sim$7\% higher energy at 5.8\,MeV instead of 5.4\,MeV. This shift is much larger than nonlinearities in the $\gamma$ region would suggest, but we note that a similar difference in the detector response for $\alpha$ particles has been observed previously with lithium molybdate based detectors \cite{Armengaud:2017,Poda:2017a}. 
Events at higher light yield than \ga/\be events can be observed due to noise spikes and misreconstructed amplitude estimates in the LD, as well as due to close \be contaminations with a coincident \ga depositing energy in the \enrLMO crystal.

We estimate a scintillation light quenching of \al-particles with respect to \ga/\be particles of $(19.7\pm1.0)$\% across the detectors (see Table \ref{tab:LMOperformance}). These results are also within expectations for this scintillation material  \cite{Armengaud:2017,Poda:2017a,Poda:2017}.

\subsection{Extrapolated $\alpha$ discrimination of \enrLMO scintillating bolometers}
\label{sec:AlphaDiscrimination}

We systematically evaluate the $\alpha$ discrimination level following Refs.~\cite{Bekker:2016,Armengaud:2017,Poda:2017a} and report the discrimination of \al versus \ga/\be events in terms of the discrimination power (DP) at the $Q$-value for \onbb in \Mo

\begin{equation}
    DP = \frac{\mu_{\gamma} - \mu_{\alpha} }{\sqrt{\sigma_{\gamma}^{2} + \sigma_{\alpha}^{2}}}.
\end{equation}

The parameters in the definition of the DP are the mean \rly{}s $\mu_{\alpha}$, $\mu_{\gamma}$ for $\alpha$ and \ga/\be events respectively, and resolutions $\sigma_{\alpha}$, $\sigma_{\gamma}$.
We obtain detector based values $\mu_{\alpha} = QF_{\alpha} \cdot \mu_{\gamma}$ from the measured $\mu_{\gamma}$ and approximate the very uniform light quenching of \al events with $QF_{\alpha} = 0.2$ (see Table \ref{tab:LMOperformance}). 
The expected LD resolutions $\sigma_{\alpha}$ and $\sigma_{\gamma}$ at the endpoint of the \Mo decay are extrapolated by adding the baseline resolution and a statistical photon noise component with an average photon energy of 2.07~\,eV \cite{Bekker:2016} in quadrature. 
The resulting median discrimination power is 15.0, with the worst-performing detector having a discrimination power of 6.3. Hence all detectors are expected to achieve better than 99.9\% $\alpha$ rejection with more than 99.9\% $\gamma$/$\beta$ acceptance.  

We note that this model calculation does not take into account additional sources of uncertainty such as variation associated with the position of the incident particle interaction and subsequent light propagation. 
However, the validity of the model is supported by the excellent agreement between the predicted and achieved discrimination in neutron calibration data in previous measurements \cite{Poda:2017a}. 
The computed discrimination level exceeds the requirements for \cupid, and we plan to study adverse effects due to non-Gaussian tails with larger statistics in the future. If multiple alpha peaks emerge in individual detectors we will also be able to study the $\alpha$ energy scale and the energy dependence of the $\alpha$ discrimination from data. It should be noted that we are only using a single of the two LDs, typically the one at the bottom of each detector module (see sec. \ref{sec:LMOperformance}). An optimized selection of the better performing LD, or a combined light estimate using both LDs will further improve the quoted discrimination. In addition it is expected that information from the combination of the LDs could potentially be relevant to break degenerecies if non-gaussian tails related to contamination at the \ntd{}s or the LDs were encountered.

\subsection{Radiopurity of \enrLMO crystals}
\label{sec:Radiopurity}

\begin{figure}[htbp]
\centering
\includegraphics[width=0.49\textwidth]{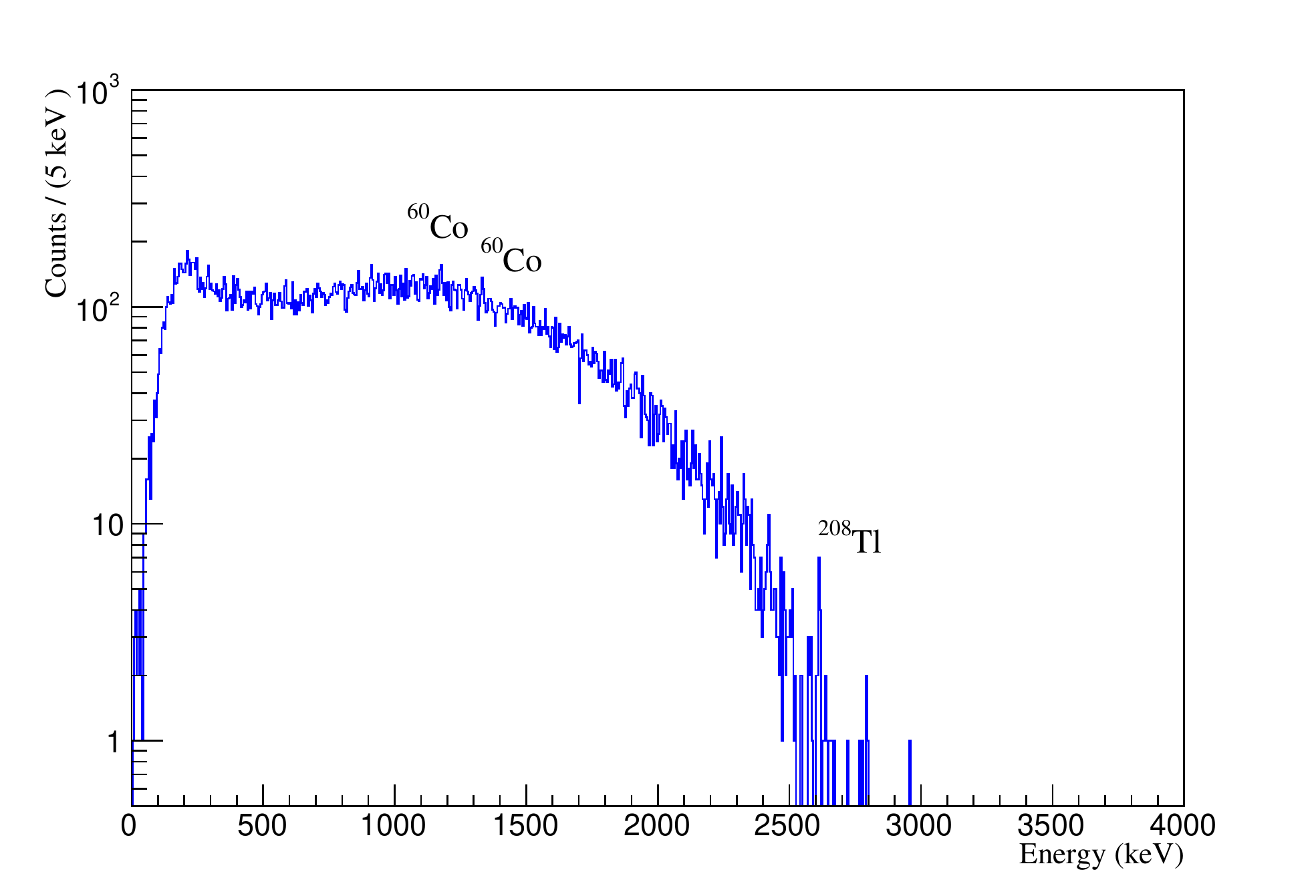}
\caption{Summed background spectrum using 11.1 days of data from 19/20 \enrLMO cryogenic detectors of the CUPID-Mo experiment.}
\label{fig:BgSpectrumGamma}
\end{figure}

We apply an additional anticoincidence cut with a time coincidence window of 100\,ms between \enrLMO detectors, a so called multiplicity one (M1) cut to reject multi-Compton and muon shower events and obtain the background spectrum shown in Fig. \ref{fig:BgSpectrumGamma}. The \ga/\be spectrum of \enrLMO bolometers above $\sim$1 MeV is dominated by the $2\nu2\beta$ decay of \Mo with an activity of 10 mBq/kg~\cite{Armengaud:2017}. 
In 11.1 days of background data, we observe no event compatible with the \rly of \ga/\be events above 3034\,keV, the $Q$-value for double-beta decay in \Mo.
The estimate for the resolution at 2615\,keV, $(6.5 \pm 1.4)$\,keV, is compatible with the prediction from the calibration data, albeit this is subject to considerable uncertainty due to the limited statistics.

\begin{figure}[htbp]
\centering
\includegraphics[width=0.49\textwidth]{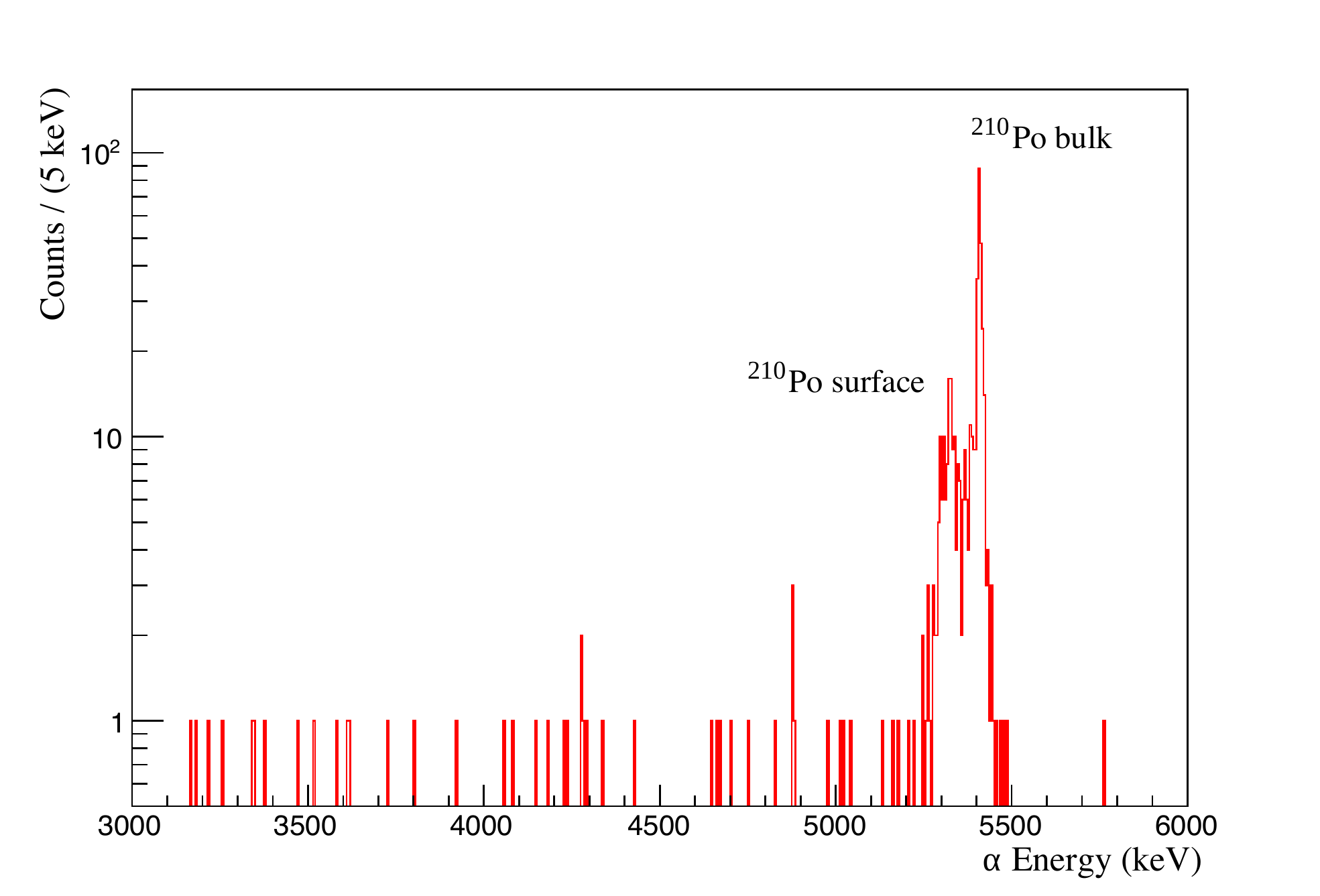}
\caption{Summed alpha region using 19/20 detectors. No RLY cut applied. (For details see text.) }
\label{fig:BgSpectrumAlpha}
\end{figure}

By relaxing the pulse shape cuts, and removing the \ga/\be RLY cuts and the M1-cut, we can investigate the \al region with a set of basic cuts that are designed to have better than 99\% acceptance.
The only clear contaminants seen in the spectrum (see Fig.~\ref{fig:BgSpectrumAlpha}) are a bulk and surface peak by \PO ($^{210}$Pb), as observed in the LUMINEU studies \cite{Armengaud:2017,Poda:2017a}. For a few of the nuclei in the Th- ($^{232}$Th, $^{228}$Th, $^{224}$Ra, $^{212}$Bi ) and the U-chain ($^{238}$U, $^{234}$U+$^{226}$Rn, $^{230}$Th, $^{222}$Rn, $^{218}$Po), that have a clean decay signature we start to see first hints of a contamination.  However, the number of events in a $\pm 30$\,keV window around the nominal decay energy is compatible with zero at the 2$\sigma$ level for all of the decay signatures. We thus place a conservative upper limit at a level of 2 \,$\mu$Bq/kg (Th-series) and 3\,$\mu$Bq/kg (U-series) (90\% C.L.) on the activity in the U-/Th-chains using the largest observed event count for any of the decays in the U-/Th-chain. 

We look for possible backgrounds from surface contaminants in the 3--4\,MeV region. 
Excluding a potential \PT alpha bulk contribution in a $\pm$30 keV window around 3269 keV we observe 14 events, which is equivalent to a background of ($0.14 \pm 0.04 $) \ckky in degraded alpha events before the rejection by \rly.

\section{Outlook}
\label{sec:outlook}

\begin{figure}[htbp]
  \centering
  \includegraphics[width=0.49\textwidth]{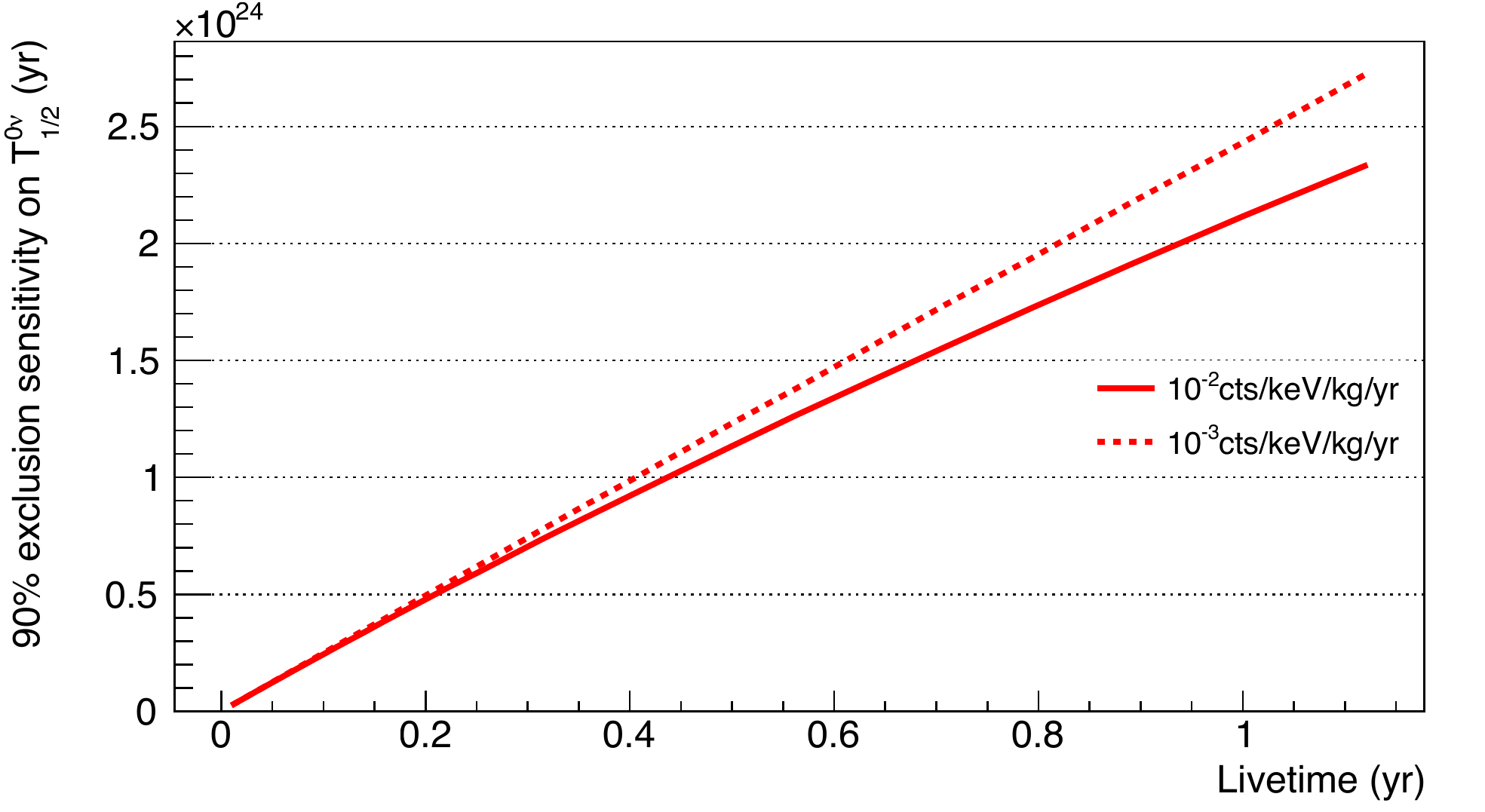}
  \caption{Bayesian exclusion sensitivity at 90\%\,C.I. for 5\,keV resolution (FWHM) at \Qbb and different background levels.}
  \label{fig:sensitivity}
\end{figure}

Based on these first physics data we are confident that the \enrLMO cryogenic detectors possess a high degree of reproducibility and are well suited to scale to a much larger CUORE sized detector array in CUPID \cite{CUPIDInterestGroup:2019inu}.
We expect the current \cupidmo experiment to be able to set significant limits on \onbb in \Mo. Consequently, we evaluate the \cupidmo sensitivity using the Bayesian method for limit setting for a counting experiment in a $\pm2\sigma$ region of interest around \Qbb. At present, several steps of the data analysis procedure have not been fully optimized, leaving room for improvement. We expect to achieve a 5\,keV energy resolution (FWHM) at \Qbb with a dedicated optimization of the energy reconstruction algorithms.
With an average containment efficiency for \onbb decay events of 75\%, and assuming a $\sim$ 90\% analysis efficiency, as obtained in CUORE \cite{Alduino:2018} and CUPID-0 \cite{Azzolini:2019tta} for the combined trigger efficiency, multiplicity, pulse shape analysis (PSA) and RLY cut efficiencies we obtain the exclusion sensitivity curves reported in Fig.~\ref{fig:sensitivity}.
If we demonstrate a background index of $10^{-2}$ \ckky with increased statistics,
\cupidmo reaches a sensitivity superior to the most recent limit on the \Mo half-life set by NEMO-3~\cite{Arnold:2015} in just 6\,months of accumulated livetime.
Fig.~\ref{fig:sensitivity} also reports the sensitivity for the more optimistic scenario
where the background level is $10^{-3}$ \ckky; in this case, the experiment is practically background free for a total of 1\,yr of livetime reaching a final sensitivity of $T_{1/2}^{0\nu\beta\beta} = 2.43\times10^{24}$\,yr. The exclusion sensitivity has a very minor dependence on the detector energy resolution and decreases by $\sim$10\% for a factor two  worse resolution and a background index of $10^{-2}$ \ckky.

\section{Conclusion}

The first physics data of the CUPID-Mo experiment validates and extends the previously reported bolometric performance for cryogenic \lmo crystals \cite{Armengaud:2017,Poda:2017a} on a much larger array of 20 detectors. We find that crystal growth and detector assembly can be well controlled to obtain excellent uniformity in performance and radiopurity. In particular, the summed energy resolution was 5.3 keV (6.5 keV) FWHM at 2615\,keV in calibration (physics) data of 19 out of 20 detectors. The measured light yield for \ga/\be events (0.6--0.9 keV/MeV), the quenching of the scintillation light for $\alpha$ particles (20\%) with respect to \ga/\be{}s and the achieved baseline resolution of bolometric Ge light detectors (146\,eV FWHM) are compatible with full \al to \ga/\be separation (median discrimination power value of 15). The \enrLMO crystals also exhibit a high level of radiopurity, particularly $\leq$3~$\mu$Bq/kg of $^{226}$Ra and $\leq$2~$\mu$Bq/kg of $^{232}$Th. 
The results indicate the prospect to surpass the sensitivity of NEMO-3 with $\sim$6 months of physics data in the current demonstrator. The technology is scalable, and the first results presented in this article strengthen the choice of \enrLMO as the baseline option for application in the CUPID next-generation cryogenic \onbb experiment. 
Additional data from the current and future demonstrators is essential to develop a detailed background model, investigate and optimize the performance in the region of interest for \onbb, and to further strengthen the projections for CUPID.

\begin{acknowledgements}

This work has been partially performed in the framework of the LUMINEU program, a project funded by the Agence Nationale de la Recherche (ANR, France). The help of the technical staff of the Laboratoire Souterrain de Modane and of the other participant laboratories is gratefully acknowledged. 
We thank the mechanical workshops of CEA/SPEC for their valuable contribution in the detector conception and of LAL for the detector holders fabrication.
The group from Institute for Nuclear Research (Kyiv, Ukraine) was supported in part by the IDEATE
International Associated Laboratory (LIA). A.S. Barabash, S.I. Konovalov, I.M. Makarov, V.N. Shlegel and V.I. Umatov were supported by Russian Science Foundation (grant No. 18-12-00003). O.G. Polischuk was supported in part by the project ``Investigations of rare nuclear processes'' of the program of the
National Academy of Sciences of Ukraine ``Laboratory of young scientists''. 
The Ph.D. fellowship of H. Khalife has been partially funded by the P2IO LabEx (ANR-10-LABX-0038) managed by the ANR (France) in the framework of the 2017 P2IO Doctoral call.
C. Rusconi is supported by the National Science Foundation Grant NSF-PHY-1614611.
This material is also based upon work supported by the US Department of Energy (DOE) Office of Science under Contract No. DE-AC02-05CH11231; by the DOE Office of Science, Office of Nuclear Physics under Contract Nos. DE-FG02-08ER41551 and DE-SC0011091; by the France-Berkeley Fund, the MISTI-France fund and  by the Chateau-briand Fellowship of the Office for Science \& Technology of the Embassy of France in the United States. This research used resources of the National Energy Research Scientific Computing Center (NERSC) and of the Savio computational cluster provided by the Berkeley Research Computing program at the University of California, Berkeley (supported by the UC Berkeley Chancellor, Vice Chancellor for Research, and Chief Information Officer).
This work makes use of the DIANA data analysis software which has been developed by the CUORICINO, CUORE, LUCIFER, and CUPID-0 Collaborations.

\end{acknowledgements}




\end{document}